\documentclass[aps,prd,superscriptaddress,showkeys,showpacs,twocolumn,a4paper]{revtex4-1}
\usepackage{ulem}
\usepackage{color}
\usepackage{graphicx}
\usepackage{soul}
\usepackage{appendix}
\usepackage{epsfig}
\usepackage[usenames,dvipsnames,svgnames]{xcolor}
\definecolor{purple}{rgb}{0.3,0,0.9} 
\definecolor{blue1}{rgb}{8, 24, 168}
\definecolor{darkteal}{HTML}{045D5D}
\usepackage[colorlinks=true,linkcolor= blue, citecolor = blue, urlcolor = purple ]{hyperref}
\usepackage{epstopdf}
\usepackage{amsmath,amssymb,amsthm,amsfonts} 
\usepackage{subfig}
\usepackage{comment}
\usepackage{float}
\UseRawInputEncoding

\usepackage{float}

\newcommand{\be}{\begin{equation}}
\newcommand{\ee}{\end{equation}}
\newcommand{\ba}{\begin{eqnarray}}
\newcommand{\ea}{\end{eqnarray}}
\newcommand{\nn}{\nonumber\\}

\begin{document}

\title{
Anomalous diffusion of the heavy quarks through the fractional Langevin equation}

\author{Jai Prakash}
\email{jaiprakashaggrawal2@gmail.com}

\affiliation{Department of Physics, Indian Institute of Technology Bombay, Mumbai 400076, India}
\affiliation{School of Physical Sciences, Indian Institute of Technology Goa, Ponda-403401, Goa, India}


\begin{abstract}

 The dynamics of heavy quarks within the hot QCD medium have been revisited, considering the influence of anomalous diffusion.  This study has been conducted using the framework of the fractional Langevin equation involving the Caputo fractional derivative. We introduce a numerical scheme for the fractional Langevin equation and demonstrate that the mean square displacement of the particle exhibits anomalous diffusion, deviating from a linear relationship with time. Our analysis calculates various entities, such as mean squared momentum, momentum spread, and the nuclear suppression factor, $R_{AA}$. Notably, our findings indicate that superdiffusion strongly suppresses the $R_{AA}$ compared to normal diffusion in the hot QCD medium. The possible impacts on other parameters are also discussed.

\end{abstract}



\maketitle

 \section{Introduction}
 \label{intro}
 
The ultra-relativistic heavy-ion collisions (HICs) at the Relativistic Heavy-Ion Collider (RHIC) and the Large Hadron Collider (LHC) have predicted the existence of Quark-Gluon Plasma (QGP); a state of matter where quarks and gluons are free to move beyond the nucleonic volume \cite{STAR:2006vcp, Adams:2005dq, PHENIX:2004vcz, ALICE:2010khr, BRAHMS:2005gow}.  The QGP is short-lived, with an expected lifespan of a few fm/c (approximately 4-5 fm/c at RHIC and 10-12 fm/c at LHC ~\cite{vanHees:2004gq, Rapp:2009my}). Conversely, investigating the characteristics of the QGP by studying the dynamics of heavy quarks (charm and beauty) is a subject of significant interest. { In this context, }The heavy quarks (HQs) serve as prominent probes of QGP \cite{Song:2015sfa, Andronic:2015wma, Dong:2019unq, Cao:2018ews, Rapp:2018qla, Prino:2016cni, Aarts:2016hap, Uphoff:2011ad, GolamMustafa:1997id, Plumari:2017ntm, Gossiaux:2008jv, Prakash:2021lwt, Prakash:2023wbs, Prakash:2023hfj,  Singh:2023smw, Kurian:2020orp, Cao:2016gvr, Mazumder:2011nj, Zhang:2022fum, PhysRevD.103.054030, Jamal:2021btg,Jamal:2020emj, Sun:2023adv,Plumari:2019hzp, Prakash_2024,Debnath:2023zet, du2023accelerated,Shaikh:2021lka,Kumar:2021goi,das2022dynamics}. Due to their large masses ($M_c/M_b$ $\sim$ 1.3/4.5 GeV), HQs are generated at the early stages of HICs. Moreover, the thermalization of the HQs is delayed compared to the light partons of the bulk medium by a factor on the order of $\sim$ $M/T$. This delay renders the thermalization time of the HQs comparable to the lifetime of the QGP fireball. 
Since the HQs are not expected to achieve complete thermalization, they retain the memory of their interaction history. Hence, they serve as a novel probe for observing the complete evolution of the QGP medium and act as non-equilibrium entities within the equilibrated QGP. 
The standard Langevin equation (LE) is used to study Brownian motion under the assumption of normal diffusion, which means a linear increase in the mean-square displacement of the particle with time for a sufficiently large time \cite{kubo2012statistical}.  By using the Langevin equation \cite{PhysRevC.86.034905, PhysRevC.93.014901, PhysRevC.84.064902,vanHees:2007me, Das:2013kea} in the context of the HQs momentum and position evolution in the hot QCD medium and numerous studies available for the experimental observables related to the HQs, such as the nuclear suppression factor,  $R_{AA}$ and elliptic flow, $v_2$ \cite{He:2022ywp, PhysRevC.86.014903, Das:2015ana, Zhang:2022fum}.
{However, the conventional approach faces challenges in adequately explaining the experimental observations concerning the energy loss experienced by HQs within the QGP medium \cite{Das:2015ana}. This prompts the question: Could the anomalous motion of HQs offer a more suitable explanation for the observed data?
Addressing this question necessitates an investigation into whether the anomalous behaviour observed in the motion of non-relativistic Brownian particles can also occur in the motion of relativistic Brownian particles. By examining this possibility, we can determine whether anomalous motion could potentially provide insights into the discrepancies observed in the energy loss of HQs within the QGP medium.} Several processes exhibit anomalous diffusion, a 
phenomenon where the mean squared displacement of the particle does not vary linearly with time as predicted by Einstein's law. Specifically, anomalous diffusion is characterized by a non-standard time dependence of the mean squared displacement ($\langle  x(t)^2\rangle$) written as follows \cite{balescu1995anomalous,bouchaud1990anomalous,metzler2000random, PhysRevLett.85.5655, PhysRevLett.104.238102,vitali2018langevin,oliveira2019anomalous,sokolov2005diffusion,lutz2001fractional,lim2009modeling, PhysRevE.107.024105,PhysRevE.108.034113,PhysRevLett.125.240606,PhysRevE.104.024115},
\begin{align}\label{MSD}
\langle  x(t)^2\rangle \propto t^{\mu}, \  {\mu} \neq 1, 
\end{align}
where $t$ is the evolution time of the particle.
The process describes normal diffusion, corresponding to the case where  ${\mu} = 1 $. Subdiffusion occurs when ${\mu}< 1 $, whereas superdiffusion occurs when $ {\mu}> 1 $ \cite{metzler2000random}. 
 Despite the crucial role played by the LE across several fields, it fails to accurately describe certain behaviours, such as anomalous diffusion (i.e., superdiffusion and subdiffusion).
 Therefore, fractional Langevin equations (FLE) have been proposed to study anomalous diffusion \cite{li2012spectral,kobelev2000fractional,coffey2012langevin,MEGIAS2024138370}. Mainardi et al., \cite{MR1611585, mainardi2008fractional, MR1611587} introduced an FLE in their groundbreaking research. The Langevin method has primarily been applied to study anomalous diffusion using Caputo fractional derivative \cite{guo2013numerics,li2013finite,miller1993introduction}. On the other hand, there are several distinct formulations for fractional derivatives, such as Riemann-Liouville derivative \cite{carpinteri2014fractals}, Riesz derivative \cite{Agrawal_2007}, Feller derivative \cite{Feller1971}, and others. In the subsequent, we will only use the Caputo fractional derivative in our analysis to study the anomalous diffusion of the particle in the medium.

 Another valuable approach to studying anomalous diffusion involves the investigation of the fractional diffusion equation, fractional Fokker-Planck equation \cite{PhysRevLett.82.3563}, and generalized Chapman-Kolmogorov equation \cite{PhysRevE.62.6233}.
 Anomalous diffusion has been used in many fields, including molecular chemistry \cite{zumofen1994spectral}, biology \cite{PhysRevLett.95.260603}, and anomalous diffusion,  polymer transport theory \cite{Shlesinger1993}. 
 Also, when accounting for interactions between the Brownian particle and the constituent particles of the medium, the fluctuations are influenced by their prior states, a phenomenon termed as memory. The current movement is affected by previous movements via a memory kernel in the generalized Langevin equation. Such memory effects can result in anomalous diffusion for a particular memory time \cite{wang1999nonequilibrium,wang1992long, PhysRevE.53.5872, PhysRevA.45.833, PhysRevE.73.061104, PhysRevLett.105.100602, PhysRevE.107.064131, PhysRevE.107.064131} and also plays a role in the QGP \cite{PhysRevD.99.076015, PhysRevC.85.054906,Ruggieri:2022kxv,Pooja:2023gqt,PhysRevD.103.034029}.

 Recently, the transverse momentum broadening of a fast parton via superdiffusion in the QCD matter has been studied in Ref. \cite{Caucal:2021lgf}. 

 With this motivation for the applications of superdiffusion on the HQs dynamics in hot QCD matter, we use FLE, a generalized form of the LE. Unlike the LE, the FLE replaces integer-order derivatives with fractional-order derivatives, specifically of Caputo type \cite{10.1111/j.1365-246X.1967.tb02303.x}. This may be the first attempt where we present an anomalous diffusion for the HQ dynamics in the QGP medium. In this article, we specifically study the effect of superdiffusion on the HQ dynamics in the QGP medium. The subdiffusion is not considered within the scope of the current analysis.

We anticipate a key outcome in our study, specifically that in the presence of superdiffusion, there will be an increase in the energy loss experienced by the HQs within the QGP medium. This conclusion is confirmed by studying various key parameters, including the mean squared displacement, mean squared momentum, momentum spread, $dN/dp_T$, and the  $R_{AA}$ of the HQs. Notably, the latter holds particular significance in the phenomenology of the HQs within the QGP. The strong suppression observed in $R_{AA}$ is anticipated to contribute significantly to developing a substantial elliptic flow ($v_2$) for the HQs.

The article is structured as follows: Section II introduces the formalism; we analytically solve the FLE of non-relativistic heavy particles using the Laplace technique, along with presenting a numerical scheme for solving the relativistic FLE. Section III is dedicated to the presentation of our results. Lastly, Section IV provides a comprehensive summary of our conclusions.


\section{The fraction Langevin equation for Brownian particles}
\label{NONRELATIVISTIC LIMIT}
For illustrative purposes, we discuss the scenario of a heavy particle with mass $M$ undergoing one-dimensional (1-D) motion within the nonrelativistic limit. We use the FLE to describe the dynamics of this particle. In this framework, the evolution of the particle's position and momentum is characterized by fractional derivatives of orders $\beta$ and $\alpha$ \cite{kobelev2000fractional,li2012spectral}, 
\begin{align}\label{Langevin_x}
&^{C} D^\beta_{0+}x(t)=\frac{p(t)}{M},
\\
&^{C} D^\alpha_{0+}p(t) = -\gamma p(t)+\xi(t),
\label{Langevin_p}
\end{align}

 where the momentum of a particle at time $t$ is denoted as $p(t)$, its position as $x(t)$. $^{C}D_{0+}^{\alpha}$ denotes Caputo fractional derivative, $\alpha$ and $\beta$ are the fractional parameter, with $n-1 < \alpha \leq n$, and $^{C}D_{0+}^{\beta}$  with $n-1<\beta\leq n$, $n \in \mathbb{N}$ ($n$ is a natural number). 
 Brownian particle encounters two distinct forces: the dissipative force, characterized by the drag coefficient, denoted by $\gamma$, and the stochastic force, denoted as $\xi(t)$. The latter governs the random noise, commonly referred to as white Gaussian noise. White noise gives rise to a fluctuating field without memory, characterized by instantaneous decay in correlations of the white noise, often referred to as a $\delta$ correlation.
The random force satisfies certain properties, such as:
\begin{align}\label{corr}
&\langle\xi(t)\xi(t')\rangle=2\mathcal{D}\delta(t-t'), \\
 &\langle\xi(t)\rangle=0,
\end{align}
 $\mathcal{D}$ is the diffusion coefficient of the heavy particle in a medium of temperature, $T$. The drag coefficient ($\gamma$) is related to the diffusion coefficient through the Fluctuation-Dissipation Theorem (FDT) as follows:
\begin{align}\label{FDT}
\gamma = \frac{\mathcal{D}}{MT}.
\end{align}

The Caputo fractional derivative \cite{10.1111/j.1365-246X.1967.tb02303.x} is defined as, 

\begin{align}\label{Caputo}
 ^{C} D^\nu_{0+}u(t) = \frac{1}{\Gamma({n-\nu})}\int_0^t \frac{u^{(n)}(s)}{(t-s)^{1+\nu-n}}ds,   
\end{align}
 where $u^{(n)}$ denotes the $n^{th}$ derivative of $u$. The fractional derivative which corresponds to the superdiffusion, i.e., when $1< \nu \leq 2$, is given by  
\begin{align}
\label{superD}
   ^{C} D^\nu_{0+}u(t) = \frac{1}{\Gamma({2-\nu})}\int_0^t \frac{u^{(2)}(s)}{(t-s)^{\nu-1}}ds,    
\end{align}
where $\Gamma(\cdot)$ denotes the gamma function. In the following subsection, we solve analytically the FLE of non-relativistic Brownian particles.

\subsection{Analytical solutions}

Analytically, we solve the FLE governing the motion of non-relativistic Brownian particles. To achieve this, we use the Laplace technique, providing analytical expressions for $\langle p^2(t) \rangle$ and $\langle x^2(t) \rangle$. Performing the Laplace transformation of Caputo derivative for $n-1 <\alpha \leq n$ and $n-1<\beta\leq n$ \cite{podlubnyacademic},
\begin{align}\label{laplace}
 \mathcal{L}\left[^{C} D^\alpha_{0+}p(t)\right](z)=z^\alpha\widehat{p}(z)-\sum_{k=0}^{n-1}z^{\alpha-k-1}\left[p^k(0)\right].   
\end{align}
In order to calculate the solution of $p(t)$ and $x(t)$ of the heavy particle for a super-diffusive process, i.e., when $1< \alpha \leq 2$ and $1< \beta \leq 2$, one can take the Laplace transform Eqs.~(\ref{Langevin_x}), (\ref{Langevin_p}) and using Eq.~\eqref{laplace}, it can be obtained easily as 
\begin{align}
\label{laplaceofx1}&z^\beta\widehat{x}(z)-z^{\beta-1}x(t_0) -z^{\beta-2}x^\prime(t_0)= \frac{\widehat{p}(z)}{M},\\
 \label{laplaceofv}&\widehat{p}(z)= \frac{z^{\alpha-1}}{z^\alpha + \gamma}p(t_0) + \frac{z^{\alpha-2}}{z^\alpha + \gamma}p^\prime (t_0)+ \frac{\widehat{\xi}(z)}{z^\alpha + \gamma}.
  \end{align}

Now, substituting Eq.~\eqref{laplaceofv} into Eq.~\eqref{laplaceofx1}, for simplicity, we take $M= 1$ \cite{Sandev2012}, we obtain
\begin{align}\label{laplaceofx}
  \widehat{x}(z)=&\;\frac{x(t_0)}{z}+ \frac{x^\prime(t_0)}{z^2}   +\frac{z^{\alpha-\beta-1}}{z^\alpha + \gamma}p(t_0) + \frac{z^{\alpha-\beta-2}}{z^\alpha + \gamma}p^\prime(t_0)  \nn & + \frac{z^{-\beta}}{z^\alpha + \gamma}\widehat{\xi}(z).
\end{align}
Taking the inverse Laplace transform of Eqs.~\eqref{laplaceofv} and \eqref{laplaceofx}, we obtain
\begin{align}
  p(t) =  &\;E_{\alpha,1}(-\gamma t^\alpha)p(t_0)+tE_{\alpha,2}(-\gamma t^\alpha)p^\prime({t_0}) \nn &+\int_0^t (t-s)^{\alpha-1}E_{\alpha,\alpha}(-\gamma(t-s)^\alpha)\xi(s)ds,
  \label{momentum}
  \end{align}
  and
  \begin{align}\label{position}x(t)=&\;x_0 +tx^\prime_0+ t^\beta  E_{\alpha,\beta+1}(-\gamma t^\alpha)p(t_0) \nn &+ t^{\beta+1}  E_{\alpha,\beta+2}(-\gamma t^\alpha)p^\prime({t_0}) \nn & + 
 \int_0^t (t-s)^{\beta+\alpha-1}E_{\alpha,\beta+\alpha}(-\gamma(t-s)^\alpha)\xi(s)ds,
\end{align}
\begin{figure*}[htp]
		\centering
        \includegraphics[scale = .37]{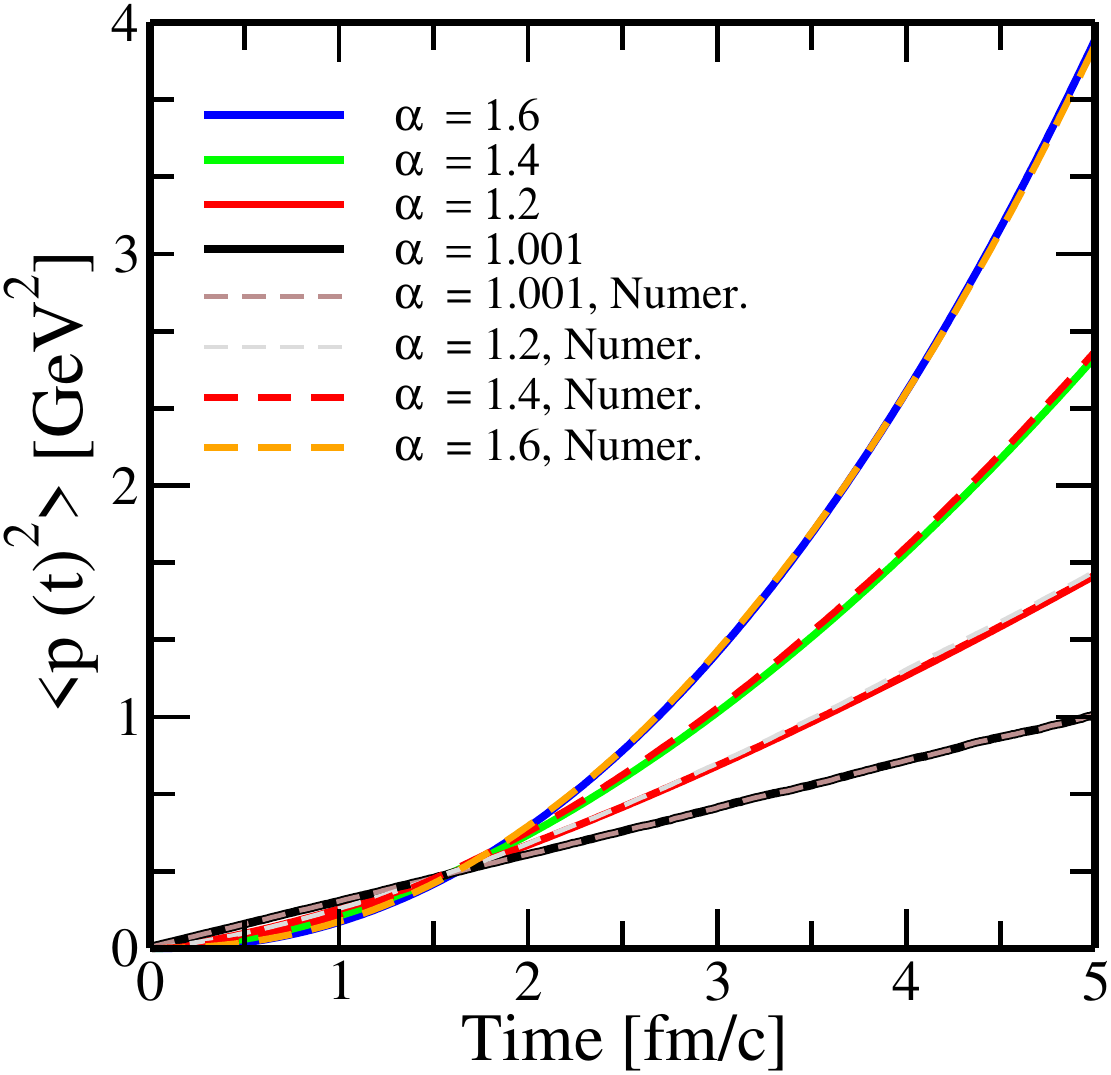}
      \hspace{10mm}
     \includegraphics[scale = .37]{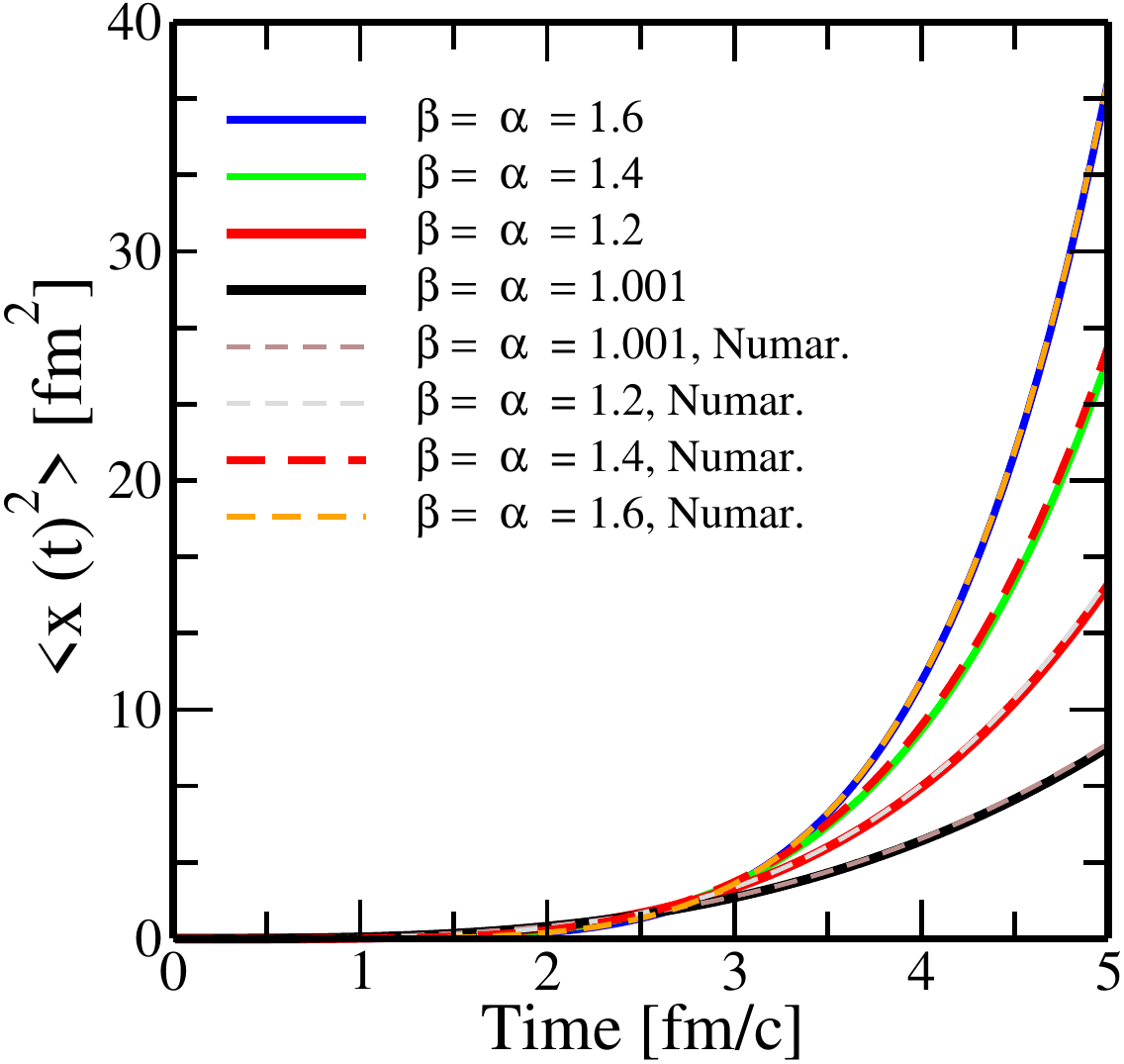}
        \caption{$\langle p^2(t) \rangle$ versus time (left panel) and $\langle x^2(t) \rangle$ versus time (right panel) for a 1-D, purely diffusive motion for different values of $\alpha$ and $\beta$. Analytic solutions are represented by solid lines, and numerical (Numer.) solutions are represented by dashed lines.}
		\label{p_1D}
	\end{figure*}
where $$
 E_{\alpha,\beta}(z) = \sum^{\infty}
_{j=0}\frac{z^j}{\Gamma(j\alpha + \beta)}$$ 
is the two-parameter Mittag-Leffler function for $\alpha>0$, $\beta>0$, and $z$ is a complex number \cite{erdelyi1953bateman} and  the Laplace transform of the two-parameter Mittag-Leffler function is \cite{MR1890104},

\begin{align}\label{Inverselaplce}
\mathcal{L}\left[x^{\beta-1}E_{\alpha,\beta}(-\lambda x^\alpha) \right](z) = \frac{z^{\alpha-\beta}}{z^\alpha + \lambda}.   
\end{align}
Next we calculate $\langle p^2(t) \rangle$, and $\langle x^2(t) \rangle$ analytically  for the Brownian particle. 

\subsubsection{Purely diffusive motion} 

We are calculating the  $\langle p^2(t) \rangle$, and $\langle x^2(t) \rangle$  for a  purely diffusive (when $\gamma =0$ in the FLE), 1-D motion. Then Eq.~\eqref{Langevin_p} simplified to
\begin{align}
^{C}D^\alpha_{0+}p(t) = \xi(t),
\label{Langevin_p_diff}
\end{align}

with initial momentum, $p(t_0)=0$. Then to calculate $\langle p^2(t) \rangle$ we follows Eq.~\eqref{momentum} and Eq.~\eqref{Langevin_p_diff}, then we get

\begin{align}\label{pt2_1D}
\langle p^2(t) \rangle= \frac{1}{\Gamma(\alpha)^2} \int_0^{t} \int_0^{t} (t-s_1)^{\alpha -1}  (t-s_2)^{\alpha -1} &\langle \xi(s_1) \xi(s_2) \rangle \nonumber \\ & ds_1 ds_2.
\end{align}
Solving further Eq.~\eqref{pt2_1D}, using the correlation properties defined in Eq.~\eqref{corr}, we have,

\begin{align}
\langle p^2(t) \rangle& = \frac{2\mathcal{D}}{\Gamma(\alpha)^2} \int_0^t (t-s)^{2\alpha -2}  ds,
\end{align}
\begin{align}
\label{momentum_analytic}
\langle p^2(t) \rangle& =\frac{2\mathcal{D}}{\Gamma(\alpha)^2} \frac{t^{2\alpha-1}}{2\alpha-1}.
\end{align}
Similarly, one can calculate analytical solutions of  $\langle x^2(t) \rangle$ for $\gamma = 0 $, and with initial conditions, $x(t_0)=0$, using Eqs.~\eqref{position}, ~\eqref{Langevin_p_diff} as

\begin{align}
   \langle x^2(t)  \rangle \approx &\; \frac{2\mathcal{D}}{\Gamma(\alpha+\beta)^2} \int_0^{t}  (t-s)^{2\alpha+2\beta -2}ds  \nonumber \\ &=\frac{2\mathcal{D}}{\Gamma(\alpha+\beta)^2} \frac{t^{2\alpha+2\beta-1}}{2\alpha+2\beta-1}.
   \label{position_analytic}
\end{align}

 We have plotted Fig. \ref{p_1D} for
the case of purely diffusive motion, we depict the variation of $\langle p^2(t) \rangle$ over time (left panel)  and $\langle x^2(t) \rangle$ (right panel), each for different values of $\alpha$ and $\beta$, while keeping $\mathcal{D}$ constant at $0.1$ GeV$^2$/fm. For simplicity, a non-relativistic case has been considered with $M =  1$ GeV. In Fig. \ref{p_1D} (left panel), results obtained from  Eq.~\eqref{momentum_analytic} show that  $\langle p^2(t) \rangle$ exhibits nearly linear evolution with time when $\alpha \rightarrow 1$ (as shown by black solid lines), which is giving $\langle p^2(t) \rangle = 2\mathcal{D}t$. In this scenario, the Caputo fractional derivative reduces to an ordinary first-order derivative; in this limit of $\alpha$, the FLE is reduced back to the LE. On the other hand, when the value of $\alpha >1$, the linear increase in $\langle p^2(t) \rangle$ with time converts into nonlinear growth over time. 
It becomes evident that the diffusion process evolves more gradually during the initial time, particularly up to $1.5$ fm/c. In contrast, around $2$ fm/c, the pattern is reversed, now a higher value of $\alpha$, giving a faster diffusion. Similarly, with the same input parameter, Fig. \ref{p_1D} (right panel) results obtained from Eq. \eqref{position_analytic} depicts the mean square displacement, $\langle x^2(t) \rangle$ varies with time. In the limit, $\alpha, \beta \rightarrow 1 $(as shown by black solid lines), $\langle x^2(t) \rangle$ varies with $t^3$ (as shown by black solid lines), for the normal diffusion. {   In this case, we considered the scenario of purely diffusive motion ($\gamma = 0$ in the FLE), where the diffusion of the Brownian particle predominates due to this approximation. Consequently, $\langle x(t)^2\rangle$ of the non-relativistic Brownian particle varies with $t^3$  for pure normal diffusion. It is crucial to emphasize that because of $\gamma$= 0, the variation of $\langle x(t)^2\rangle$ with $t^3$ is unrelated to  Eq~\eqref{MSD}. 
Given the limiting nature of this scenario, the calculated results seem to contradict the expected behaviour of normal diffusion, as expected in Eq. \eqref{MSD}. Once our analytical and numerical calculations align, we will incorporate the drag term into the relativistic Langevin equation for the charm quark in the subsequent subsection. The reason for choosing this limiting case will be discussed in the following subsection.}
With the higher values of $ \alpha$ and  $\beta$,  $\langle x^2(t) \rangle$ varies with the power of time greater than cubic power.

\subsection{ A numerical scheme for the fractional Langevin equation}

In classical stochastic differential equations driven by Brownian motion, the It\^o or Stratonovich stochastic calculus is typically employed for solutions. However, these methods are not applicable to the FLE driven by fractional Brownian motion, as it is not a semimartingale (see Ref. \cite{rogers1997arbitrage} for detailed proof). Although the Monte Carlo method is a reliable solution for stochastic differential equations, it is not appropriate in this case because it relies on independent sequences. Still, the sequences in fractional noise are dependent. Therefore, we have employed L2 numerical schemes, as detailed in \cite{MR361633}. These schemes are among the most effective for discretizing the Caputo fractional derivative and provide a key contribution to this article.
We numerically solve the FLE in Eqs.~\eqref{Langevin_x}, \eqref{Langevin_p}. To validate our numerical computations, we compared the numerical results with the analytical ones for the 1-D FLE, considering $\gamma$ = 0. This step is crucial to ensure the accuracy of our numerical method, which will be employed in solving the relativistic Langevin equation that will be discussed in the following subsection.
To solve the FLE, numerical methods are commonly categorized as indirect or direct. Since time-fractional differential equations can generally be reformulated into integro-differential equations, solving such equations corresponds to indirect methods. On the other hand, direct methods focus on approximating the time-fractional derivative itself. This aspect constitutes the core of our work, where our goal is to discretize the fractional derivative directly without transforming the associated differential equation into its integral form. Our numerical algorithm for solving the FLE is based on the three-step scheme using the central difference method shown in Appendix \ref{AP}.

\subsubsection{The fractional Langevin equation of relativistic Brownian particles: QGP}

We extended the solution of FLE { for the non-relativistic Brownian pariticle} as defined in Eqs.~\eqref{Langevin_x},\eqref{Langevin_p}, to describe the HQs dynamics in the QGP medium in the relativistic limit as follows,

\begin{align}\label{Langevin_x_rel}
&^{C} D^\beta_{0+}x(t)=\frac{p(t)}{E(t)},
\\
&^{C} D^\alpha_{0+}p(t) = -\gamma p(t)+\xi(t),
\label{Langevin_p_rel}
\end{align}
where,  $E = \sqrt{p^2+ M^2}$ represents the energy, and $p$ denotes the momentum of the HQs. The $\gamma$ can be related to the diffusion coefficient through an FDT as \cite{PhysRevLett.84.31, Moore:2004tg,Mazumder:2013oaa}, 
\begin{align}\label{FDT_rel}
\gamma = \frac{\mathcal{D}}{ET}.
\end{align}
Since both Eq.~\eqref{Langevin_x_rel} and Eq.~\eqref{Langevin_p_rel} are non-linear differential integral equations, preventing analytical solutions through methods such as Laplace transformations, as done for the non-relativistic case for Eqs.~\eqref{Langevin_x},\eqref{Langevin_p}. The solution of relativistic FLE defined in Eq.~\eqref{Langevin_p_rel} and  Eq.~\eqref{Langevin_x_rel} is possible solely through numerical methods. We utilize the numerical approach to compute $\langle x^2(t) \rangle$, $\langle p^2(t) \rangle$, the momentum distribution, and $R_{AA}$ of the  HQs. For superdiffusion, the relativistic FLE  can be written in the discrete form using Eq.~\eqref{SuperA} ( given in appendix \ref{AP}) as follows,

\begin{widetext}
\begin{align}\label{SuperA_x}
\begin{cases}
  x(t_1)= x(t_0) + \left[\displaystyle\frac{p(t_0)}{E(t_0)}\right]k  \; &:\;n=1, \\
  x(t_2)= 2x(t_1)-x(t_0) + \left[\displaystyle\frac{p(t_1)}{E(t_1)}\right]k \; &:\;n=2, \\
  x(t_n)= 2x(t_{n-1})-x(t_{n-2})+ \Bigg[\displaystyle\frac{p(t_{n-1})}{E(t_{n-1})}&\\  
 -\displaystyle\sum_{j=1}^{n-2} b_j \left(x(t_{n-j})-2x(t_{n-j-1}) + x(t_{n-j-2})\right)+\displaystyle\frac{n^{2-\alpha} - (n-1)^{2-\alpha}}{k}(x(t_1) - x(t_0))\Bigg]k \; &:\;n \geq 3, 
 \end{cases}
\end{align}
and
\begin{align}\label{SuperA_p}
\begin{cases}
 p(t_1)= p(t_0) + \left[-\gamma p(t_0)+\sqrt{2\mathcal{D}/\Delta{t}}\eta\right]k  \; &:\;n=1, \\
  p(t_2)= 2p(t_1)-p(t_0) +  \left[-\gamma p(t_1)+\sqrt{2\mathcal{D}/\Delta{t}}\eta\right]k \; &:\;n=2, \\
 p(t_n)= 2p(t_{n-1})-p(t_{n-2})+ \Bigg[-\gamma p(t_{n-1})+\sqrt{2\mathcal{D}/\Delta{t}}\eta&\\ 
 -\displaystyle\sum_{j=1}^{n-2} b_j\left(p(t_{n-j})-2p(t_{n-j-1}) + p(t_{n-j-2})\right)+\displaystyle\frac{n^{2-\alpha} - (n-1)^{2-\alpha}}{k}(p(t_1) - p(t_0)) \Bigg]k \; &:\;n \geq 3, 
 \end{cases}
\end{align}
\end{widetext}
where
$k = \Delta t^{\alpha}\Gamma(3-\alpha)$,  with $t_{1}= t_{0} + \Delta{t}$, $t_{n}= t_{n-1} +\Delta t$. Here $t_{0}$ is the initial time, and $\Delta{t}$ is the time step.

 \begin{figure*}[htp]
		\centering
        \includegraphics[scale = .37]{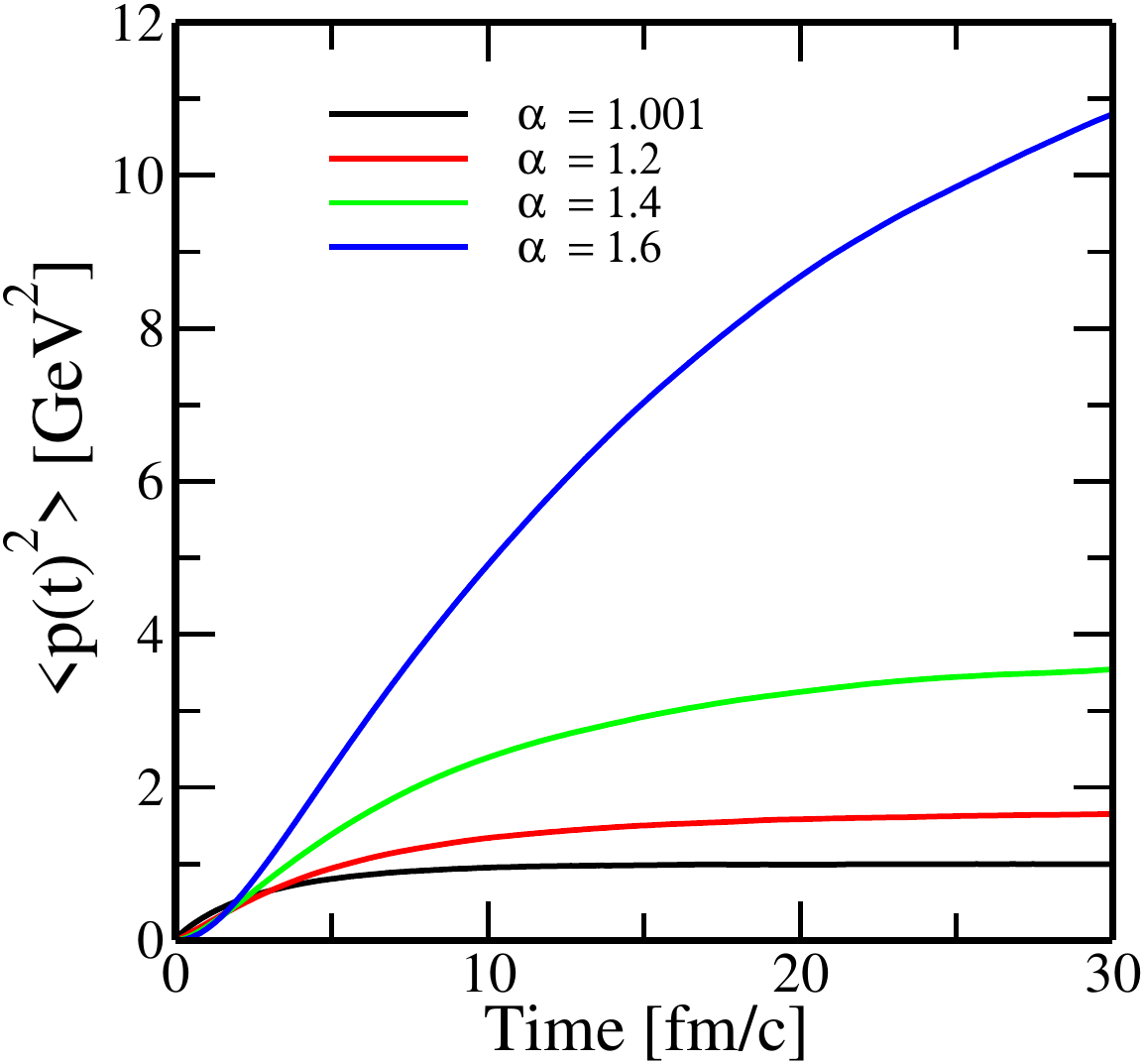}
         \hspace{10mm}
		\includegraphics[scale = .37]{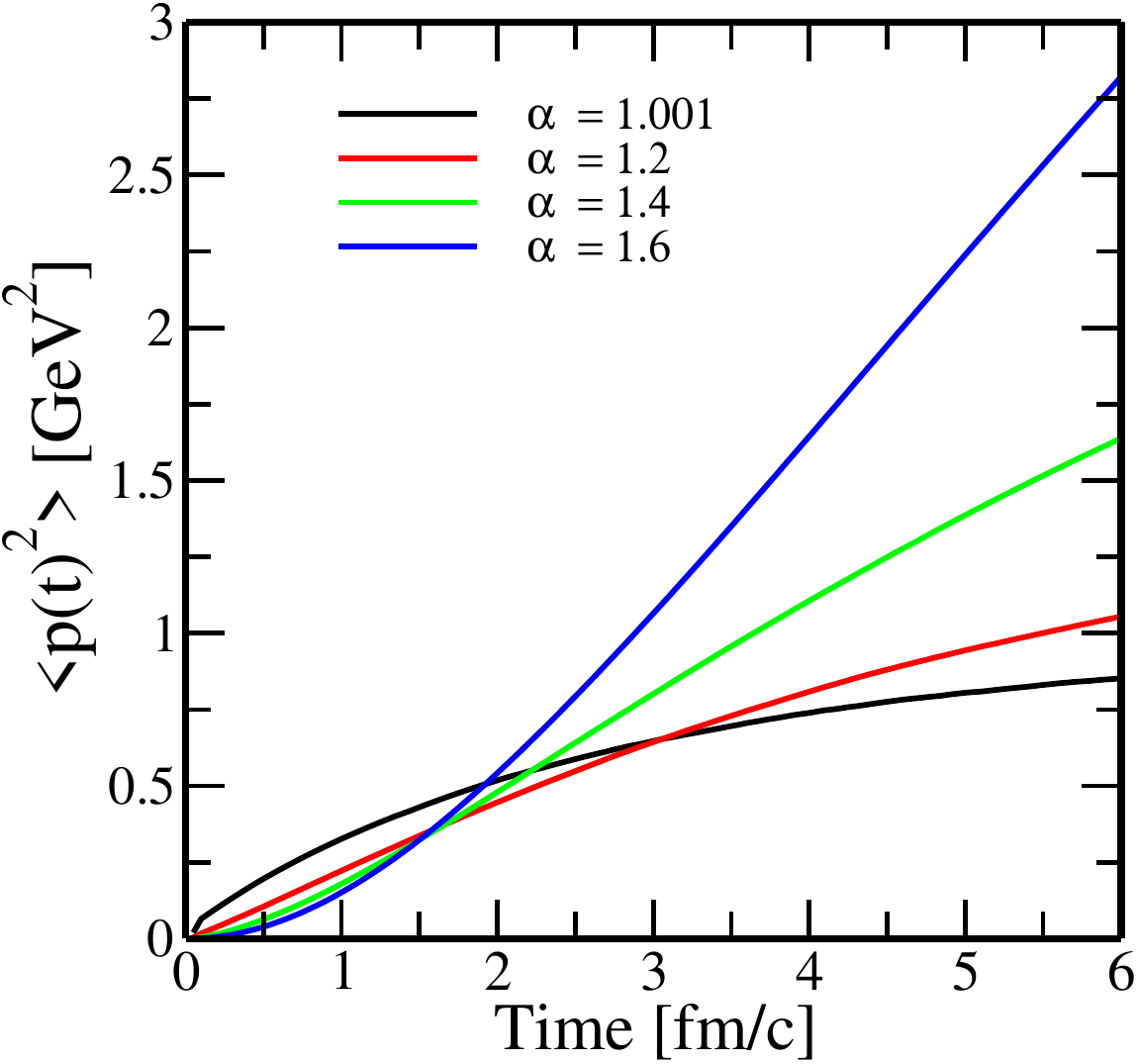}
        
		\caption{$\langle p^2(t) \rangle$ versus time, for the various values of the $\alpha$ and $\mathcal{D} = 0.1$ {GeV}$^2$/fm at $T = 250$ MeV. }

		\label{p_2D}
	\end{figure*}

\section{Results}

\subsection{ A check for non-relativistic}
To analyze the numerical scheme of FLE, we compute $\langle x^2(t) \rangle$ from Eq.~\eqref{SuperA_x} and $\langle p^2(t) \rangle$ from Eq.~\eqref{SuperA_p}, subsequently comparing the results with the analytical solutions derived from Eq.~\eqref{position_analytic} and Eq.~\eqref{momentum_analytic}, respectively. The FLE has been solved for the 1-D motion of the heavy particle within the nonrelativistic limit, as explained in Sec.~\ref{NONRELATIVISTIC LIMIT}. For illustrative purposes, we have considered constant diffusion coefficient, $\mathcal{D}$ = 0.1 GeV$^2$/fm and mass of the heavy particle, $M$ = 1 GeV and $\gamma = 0$ ({\it i.e.,} pure diffusion case) in FLE. In the left panel of Fig. \ref{p_1D}, the numerically  computed $\langle p^2(t) \rangle$ for $\alpha = 1.001$ (brown dashed line),  $\alpha = 1.2$ (grey dashed line),  $\alpha = 1.4$ (red dashed line), and  $\alpha = 1.6$ (orange dashed line) is presented. Notably, these numerical results align with the analytic solutions from Eq.~\eqref{momentum_analytic} and match the other values of the $\alpha$. For an additional test of our numerical approach, we have calculated $\langle x^2(t) \rangle$ of the heavy particle as shown in Fig. \ref{p_1D} (right panel). In the same figure, the numerically calculated $\langle x^2(t) \rangle$ for $\alpha = \beta = 1.001$ (brown dashed line) and $\alpha = \beta = 1.2$ (grey dashed line), $\alpha = \beta = 1.4$ (red dashed line), and  $\alpha = \beta = 1.6$ (orange dashed line) is displayed. Again, the numerical results match the analytic results from Eq.~\eqref{position_analytic}. One can notice that the numerical simulation agrees
with the analytical result, indicating that our numerical scheme works correctly. 

\subsection{The evolution of $\langle p^2(t)\rangle$ and $\langle x^2(t)\rangle$ of the HQs.}

For our analysis, the definition we used for $\langle p^2(t) \rangle$ is given by 
\begin{align} \langle p^2(t) \rangle = 
  \langle p_x^2(t) + p_y^2(t) \rangle. 
\end{align} 
 We have computed $\langle p^2(t) \rangle$ over time for four different values of $\alpha$, namely $\alpha = 1.001$ (black line), $\alpha = 1.2$ (red line), $\alpha = 1.4$ (green line), and $\alpha = 1.6$ (blue line) at $T = 250 $ MeV, $M$ = 1.3 GeV and $\mathcal{D} = 0.1$ GeV$^2$/fm correspond to the charm quark. The corresponding results are shown in Fig.~\ref{p_2D}. In this scenario, the initial momentum is set to be $p_x(t_0) = p_y(t_0) = 0$. 

From the left panel of Fig.~\ref{p_2D}, it is evident that as the magnitude of $\alpha$ increases, the behaviour of the process indicates superdiffusion for the charm quark within the hot QCD medium. The impact of superdiffusion becomes more noticeable at higher values of $\alpha$. 
It is noticeable that as $\alpha \rightarrow 1$, the superdiffusion process converges back to normal diffusion. Additionally, in the later stages, the mean squared momentum $\langle p^2(t)\rangle$ tends to approach $3MT$ \cite{Moore:2004tg}, and also the FLE defined in Eq.~\eqref{SuperA_p} simplifies to the standard LE described in Ref. \cite{Moore:2004tg, Das:2013kea, PhysRevC.86.014903, PhysRevC.86.034905, PhysRevC.93.014901}. This transition emphasizes the connection between superdiffusive behaviour governed by the FLE and the conventional diffusion process characterized by the standard LE. In the right panel of Fig.~\ref{p_2D}, we present a subset of the results depicted in the left panel, focusing on the early-time evolution of $\langle p^2(t) \rangle$ for four different values of $\alpha$. It is worth noting that in the initial time, the diffusion process gradually evolves for higher values of $\alpha$. However, for times beyond $3$ fm/c, the trend reverses, and larger $\alpha$ corresponds to a faster diffusion, as anticipated in Fig.~\ref{p_2D} (left panel).

\begin{figure}
		\centering
		\includegraphics[scale = .37]{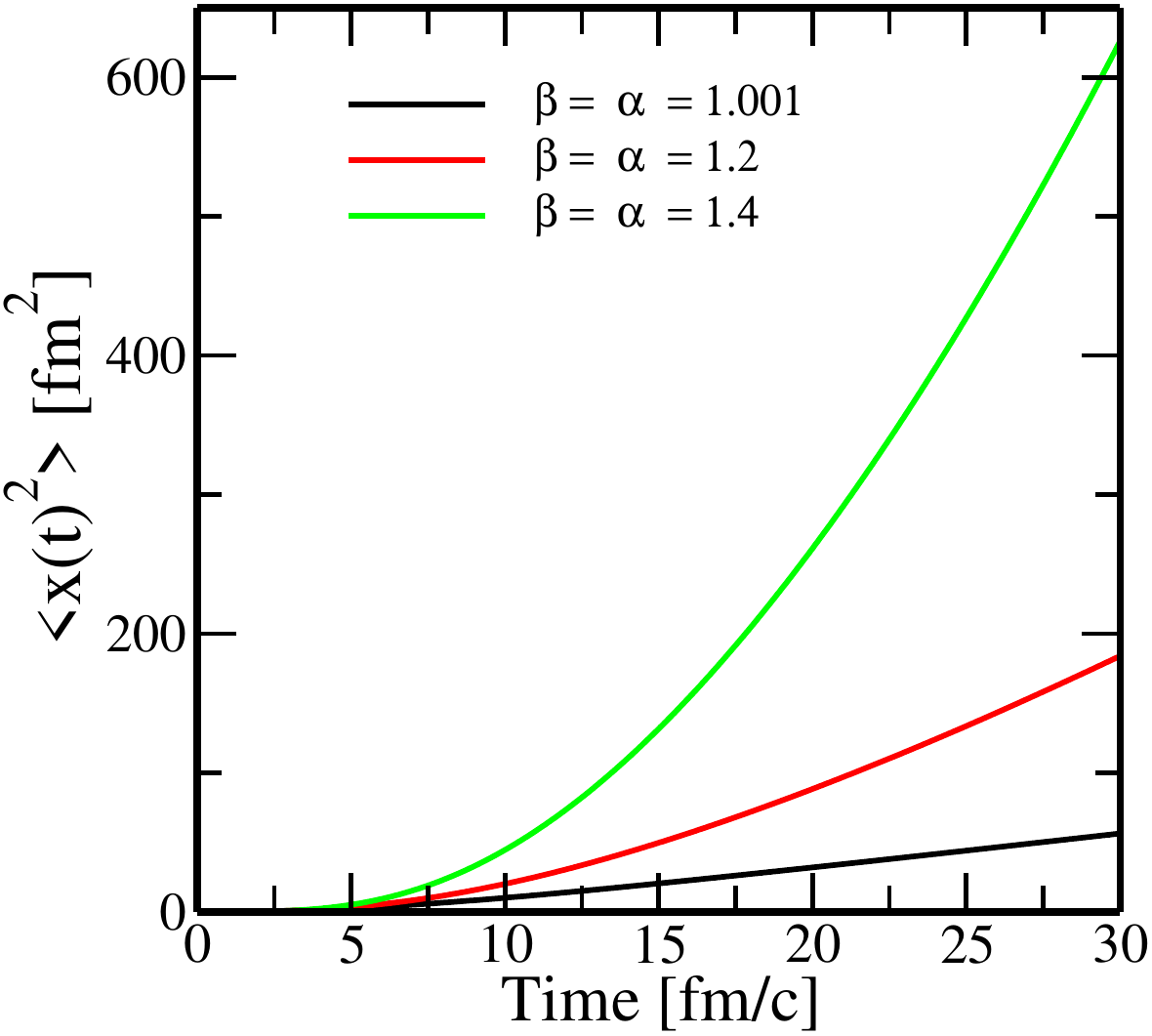}
        \caption{$\langle x^2(t) \rangle$ versus time, for the various values of the $\alpha$, $\beta$ and $\mathcal{D} = 0.1$ {GeV}$^2$/fm at $T = 250$ MeV.}
        
		\label{x_2D}
	\end{figure}

In Fig.~\ref{x_2D}, we have calculated the evolution of $\langle x^2(t) \rangle$  { of the relativistic charm quark} over time for three values of $\alpha = \beta = 1.001, 1.2, 1.4$, maintaining other parameters consistent with those illustrated in Fig.~\ref{p_2D}. The initial conditions are set to be $x(t_0) = y(t_0) = 0$. 
It can be noticed that both $\alpha = \beta =1.2, 1.4$, a distinct shift towards superdiffusion is observed (shown in Fig.~\ref{x_2D}) as we discussed initially in Eq.~\eqref{MSD}. The larger values of $\alpha$ and $\beta$ contribute to this notable change in behaviour, underlining the complex dynamics of the system. As $\alpha \rightarrow 1$ and $\beta \rightarrow 1$, the behaviour reflects normal diffusion. In this limit, at a later time, $\langle x^2(t) \rangle$ exhibits a proportional relationship with $t$ {(same as in Eq. \eqref{MSD})}, as described in \cite{Moore:2004tg, Svetitsky:1987gq}.
This puzzling behaviour of $\langle p^2(t) \rangle$ and $\langle x^2(t) \rangle$ due to the anomalous diffusion, which was not explained before in the context of the HQ dynamics in a hot QCD medium.

In the following sections, we have calculated $R_{AA}$ and the charm quark momentum distribution, $dN/dp_T$, to see the effect of superdiffusion processes.
To determine the interaction of the HQ with the thermalized bath consisting of massless quarks and gluons, we employ perturbative Quantum Chromodynamics (pQCD) transport coefficients for elastic processes with the well-established diffusion coefficients~\cite{Svetitsky:1987gq}. Here, the Debye mass, $m_{D} =  g_{T}T$ screens the infrared divergence associated with the $t$-channel diagrams.

\begin{figure*}
		\centering
        \includegraphics[scale = .3]{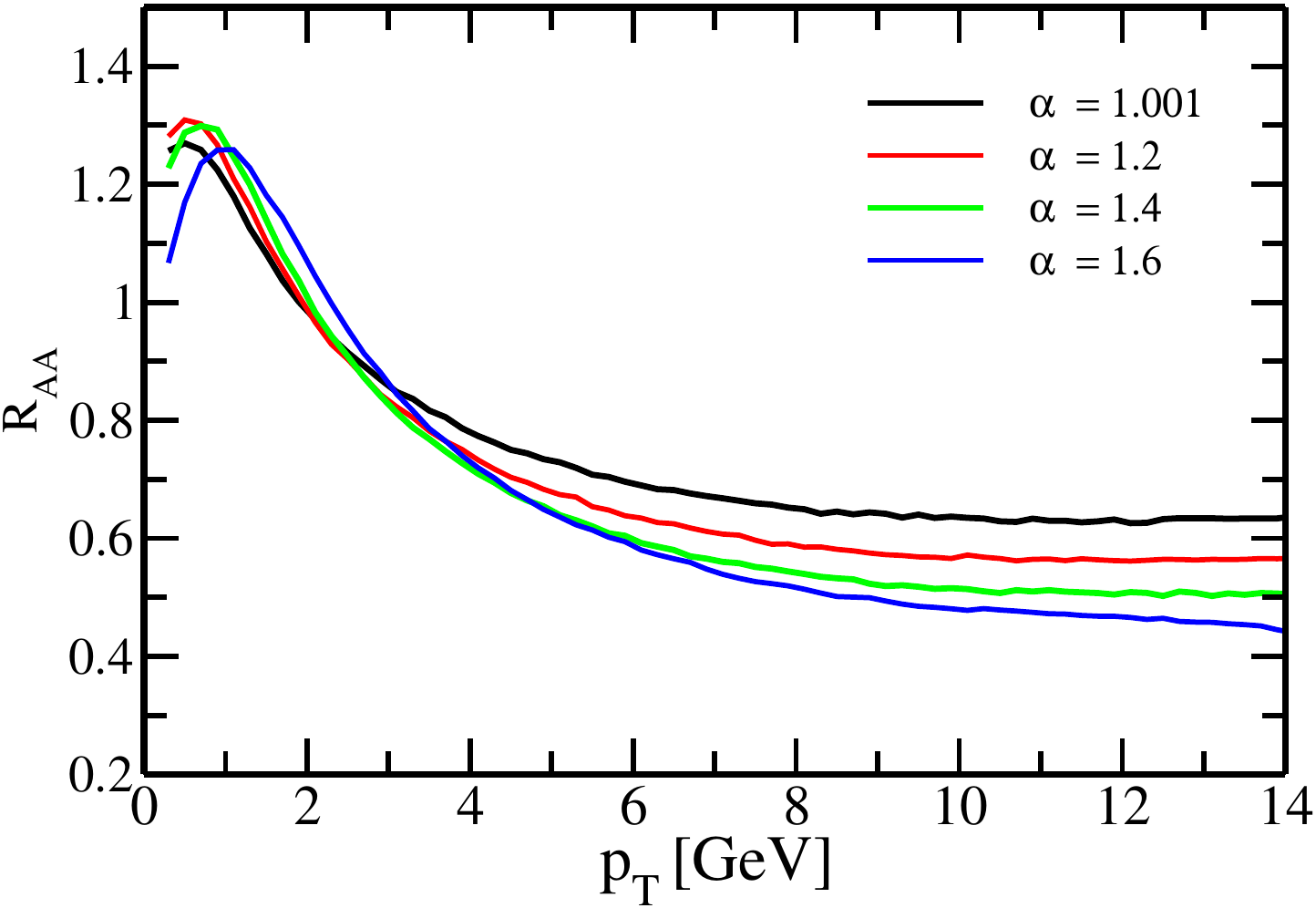}
         \hspace{10mm}
		\includegraphics[scale = .3]{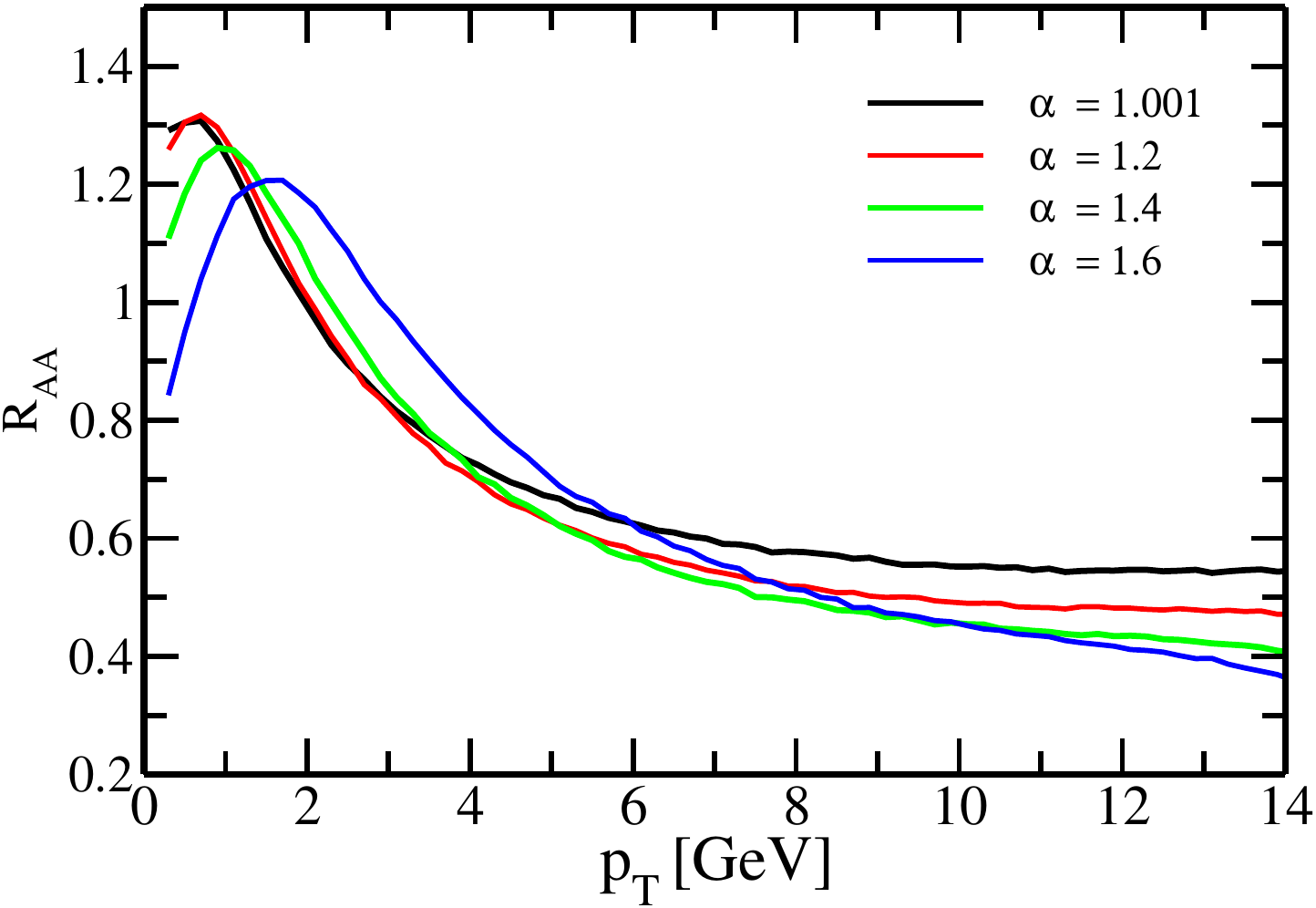}
        
		\caption{ The $R_{AA}$ as a function of $p_T$, with $t$ = 6 fm/c, and at two different temperatures ($T$ = 250 MeV (left panel) and $T$ = 350 MeV (right panel) with four different values of  $\alpha$. }

		\label{raa}
	\end{figure*} 
\subsection{Nuclear modification factor}

To analyze the impact of the superdiffusion on the experimental observable, we have calculated the nuclear suppression factor, $R_{AA}(p_T)$ for the HQs, which is defined as follows \cite{Moore:2004tg},

\begin{align}
 R_{AA}(p_T)=\frac{f_{\tau_f} (p_T )}{f_{\tau_0} (p_T)}.   \end{align}
The momentum spectrum,  $f_{\tau_f}(p)$, of charm quarks calculated for the time evolution, $\tau_f$ = $6$ fm/c in our computational results and $f_{\tau_i}(p)$ is initial momentum distribution of the charm quark. The initial momentum spectra, $f_{\tau_i}(p)$, is taken according to the fixed order + next-to-leading log (FONLL) calculations, which has been shown to be capable of reproducing the spectra of D-mesons produced in proton-proton collisions through fragmentation~\cite{cacciari2005qcd, Cacciari:2012ny},  the  initial momentum spectrum is written as,
 \begin{equation}
     \frac{dN}{d^2p{_T}} = \frac{x_0}{(x_1+p_T)^{x_2}},
\end{equation}
where the parameters are estimated as follow;  $x_0= 6.365480 \times{10}^8$, $x_1= 9.0$ and $x_2= 10.27890$. 
$R_{AA} (p_T)$ $\ne$ 1 implies that
charm quarks undergo interactions with the
medium. These interactions lead to modifications in the spectrum of charm quarks.


In Fig.~\ref{raa}, the behaviour of $R_{AA}$ is depicted as a function of $p_T$, which are calculated using the FLE as defined in Eq.~\eqref{SuperA_p}  for different values of $\alpha$.
The calculations are performed at two distinct temperatures, $T$= 250 MeV (left panel) and $T$= 350 MeV (right panel). At $T$= 250 MeV, the dominating influence is the drag force across the entire range of $p_T$. As $p_T$ increases, the significance of energy loss becomes more pronounced, which can be noticed in Fig.~\ref{raa} (left panel). When $\alpha > $ 1, the normal diffusion converts into superdiffusion, it is evident for $\alpha = 1.2$ (red line), $\alpha = 1.4$ (blue line), and $\alpha = 1.6$ (green line). With an increasing value of $\alpha$, there is a notable decrease in magnitude $R_{AA}$ (more suppression) at high $p_T$. For $\alpha \rightarrow 1$ (depicted by the black line), the behaviour of the $R_{AA}$  aligns with normal diffusion, consistent with findings available in the literature for the same input parameters \cite{PhysRevC.93.014901, Das:2013kea,Das:2015ana}. Fig.~\ref{raa} (right panel), corresponds to $T=350$ {MeV}, the observed behaviour is attributed to the diffusion-dominated propagation of the HQs within the hot QCD medium. This dominance of diffusion mechanisms effectively leads to the diffusion of low-momentum charm quarks to higher momentum states. The high temperature, $T=350$ {MeV}, enhances the significance of diffusion processes, resulting in a distinct pattern in the $R_{AA}$ as compared to $R_{AA}$ at $T$= 250 MeV. For $\alpha > $ 1, a significant reduction in the magnitude of $R_{AA}$ is observed, indicating more suppression at high $p_T$. Notably, for the highest considered value of $\alpha = 1.6$, a larger proportion of particles tends to remain at low $p_T$, a consequence of the superdiffusion process. This behaviour underscores the complex dynamics associated with superdiffusion and its impact on the $R_{AA}$ of the HQs in the QGP medium.

\subsection{Momentum spread of HQs}
\begin{figure}
		\centering
        \includegraphics[scale = .3]{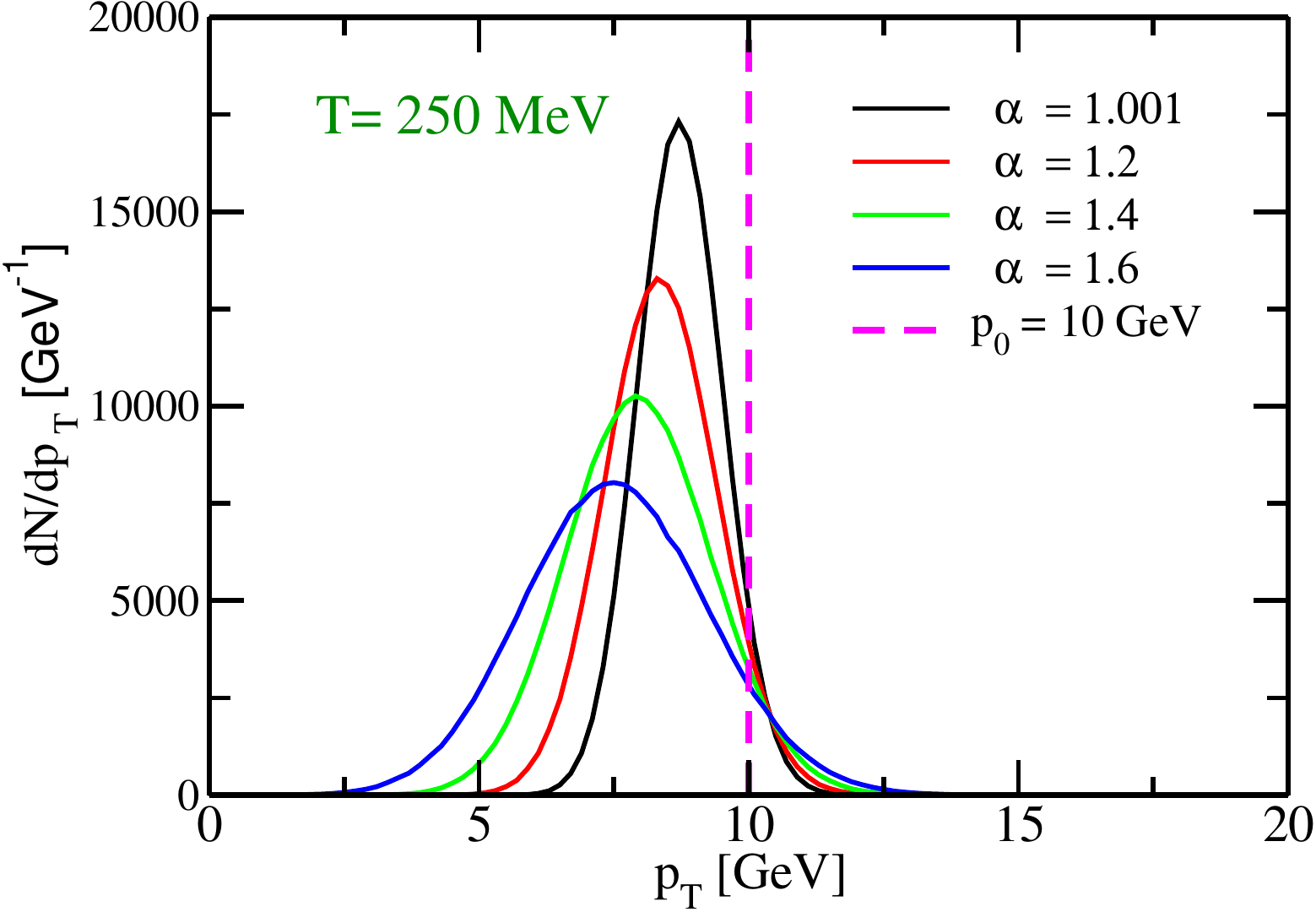}
         \hspace{10mm}
		\includegraphics[scale = .3]{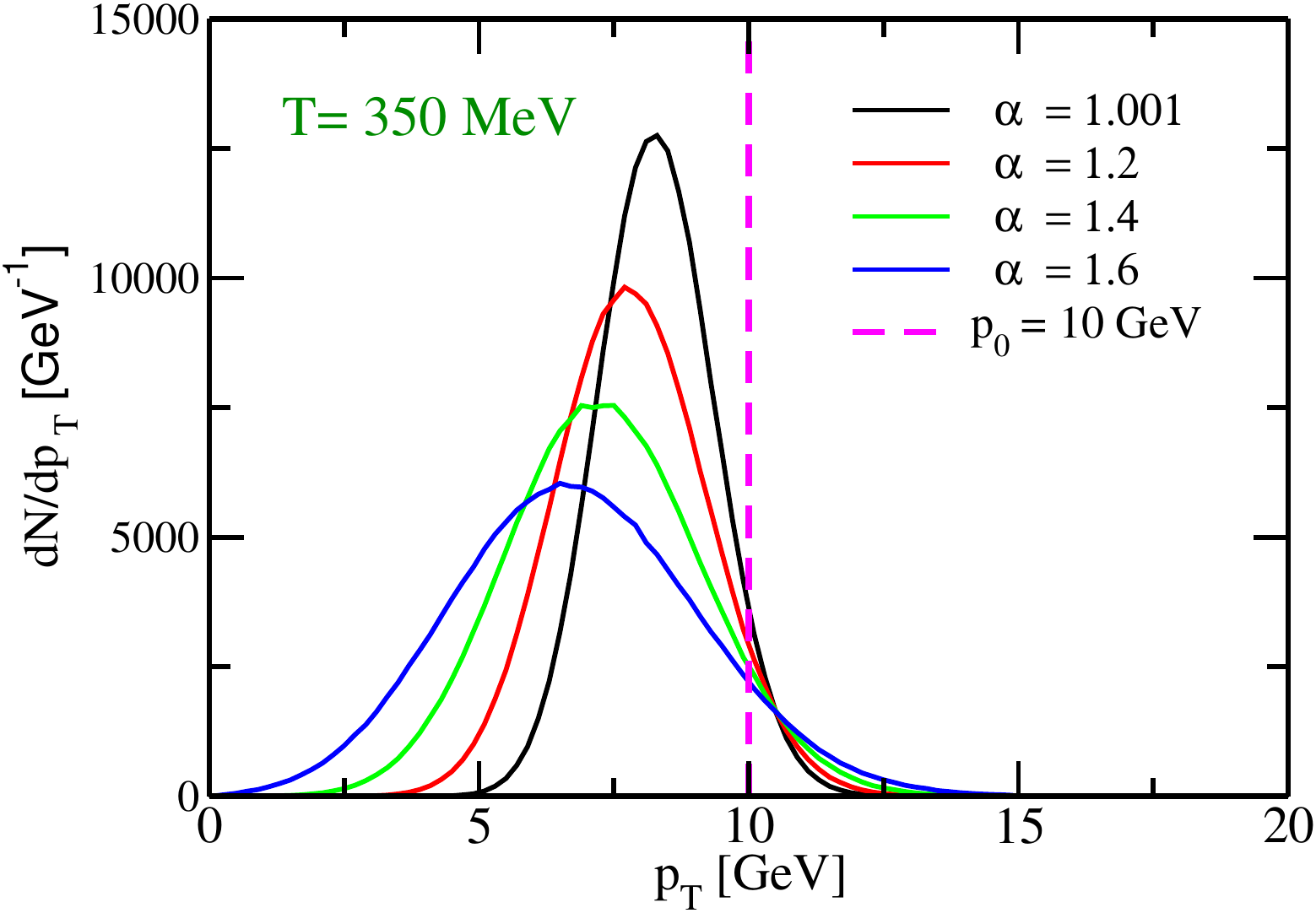}
        
		\caption{ The evolution of charm quark momentum distribution at $\tau_f$ = 6 fm/c  as a function of $p_T$, for various $\alpha$ and at two different temperatures ($T$ = 250 MeV (above panel) and $T$ = 350 MeV (below panel). Assuming an initial delta distribution centered at $p(t_0)$ = 10 GeV.}
\label{momentum_distribution}
	\end{figure}

 We show the evolution of charm momentum distribution, $dN/dp_T$, at static temperatures, $T$ = 250 MeV (top panel) and $T$ = 350 MeV (bottom panel) in Fig.~\ref{momentum_distribution}. The $dN/dp_T$ evolution is performed for various values of $\alpha$ at a final evolution time of $\tau_f$ = 6 fm/c. To understand the impact of superdiffusion on charm quarks within the QGP, we take initial conditions where all charm quarks are concentrated within an extremely narrow $p_T$ bin, creating a delta-like distribution at $p (t_0)$ = 10 GeV (magenta line). It is observed that the interaction of the HQs with the QGP medium results in the spreading of $dN/dp_T$. Subsequently, the evolution of this distribution is analyzed using the FLE as defined in Eq.~\eqref{SuperA_p}. We have observed the evolution of $dN/dp_T$  at higher values of $\alpha$ for the case of superdiffusion, where, $\alpha = 1.2$ (red line), $\alpha = 1.4$ (blue line), and $\alpha = 1.6$ (green line). As depicted in Fig.~\ref{momentum_distribution}, when $\alpha > 1$, the distribution $dN/dp_T$ undergoes a notable spread and average momentum shifts towards lower values of $p_T$ under the influence of diffusion and drag coefficients, respectively. Specifically, for $\alpha = 1.6$ (represented by the blue line), the extent of spreading is more pronounced, and the average momentum shifts towards the lower $p_T$ compared to other $\alpha$ values such as 1.001, 1.2, and 1.4.  However, for $\alpha$ = 1.001 (black line), corresponding to the normal diffusion coefficient as explained in Ref. \cite{PhysRevC.84.064902, Das:2013kea}. At the same time, the total area under the curve remains constant for all values of $\alpha$ and both temperatures.

\section{Conclusion and outlooks}

In this paper, we have discussed anomalous diffusion through the FLE with the Caputo fractional derivative, specifically focusing on superdiffusion in the context of the HQs in the QGP medium.  Notably, the mean squared displacement of the particle exhibits a power-law dependence on time (as shown in Fig.~\ref{x_2D}). Our analysis discussed the scenarios where the values of $\alpha$ and $\beta$ were taken as 1.001, 1.2, 1.4, and 1.6, showcasing the effects of superdiffusion. We have demonstrated that as $\alpha$ approaches 1, the superdiffusion reverts to normal diffusion, verified by numerical and analytical calculations for $\langle x^2(t) \rangle$ and $\langle p^2(t) \rangle$ in the non-relativistic 1-D case (see Fig.~\ref{p_1D}).
{Initially, in Section II.A.1, we focused on purely diffusive motion with $\gamma$ = 0 in the FLE. Because of the dominance of the diffusion term, $\langle x(t)^2\rangle$ varies with $t^3$, contrary to normal diffusion expectations. Upon aligning analytical and numerical results, then we later incorporated $\gamma$ into relativistic FLE, confirming normal diffusion for relativistic charm quarks (as discussed in Eq. \eqref{MSD}).} Several key quantities characterizing the dynamics of the HQs under superdiffusion have been computed, including the $\langle p^2(t)\rangle$, $\langle x^2(t)\rangle$. 
Extending our analysis, we incorporated physical observables $R_{AA}$ of the HQs in the QGP medium. We then shifted our focus to the momentum spread, $dN/dp_T$, utilizing an initial momentum distribution at $p (t_0) = 10$ GeV. The FLE, with the HQs moving under dissipative and random forces, was solved with transport coefficients serving as input parameters. Our findings indicate that superdiffusion results in more suppression in $R_{AA}$.  To select specific values for $\alpha$ and $\beta$ within the ranges of $1<\alpha\leq 2$ and $1<\beta\leq 2$, it is essential to be able to simultaneously describe the experimental observables, such as $R_{AA}$ and $v_{2}$, for the entire measured range of $p_T$. { Initially, the formation of a small $R_{AA}$ (signifying strong suppression) can occur rapidly at the beginning of the QGP, while the development of significant $v_2$ is more sensitive to later stages of evolution. Consequently, substantial interactions may not coincide with a significant build-up of $v_2$ since the bulk medium has yet to establish significant elliptic flow.
Exploring anomalous diffusion may offer the potential to generate notable $v_2$ magnitude, possibly leading to intensified interactions between the HQs and the evolving bulk at later stages. However, a thorough analysis of $v_2$ in the presence of anomalous diffusion requires a refined study incorporating realistic initial conditions, including the initial geometry and expansion of the fireball. Subsequent comparison with experimental data will be crucial for validating and refining these findings, representing an essential aspect of future investigations in this study.}

In the future, we plan to explore the impact of superdiffusion on the HQ dynamics in the QGP medium, especially considering the time correlation of thermal noise. This study devotes the groundwork for more realistic conditions, which should incorporate an exact initial geometry and an expanding medium in the near future. 
Superdiffusion might impact various observables, such as HQs directed flow, particle correlations, etc. Given the simplifying assumptions made in our current study, it is challenging to anticipate the specific modifications. A more comprehensive and quantitative analysis will be performed in near future investigations to understand these phenomena further.

\section{Acknowledgments} 

I gratefully acknowledge Dr. Santosh Kumar Das for his invaluable advice, Aditi Tomar for insightful discussions and inspirational encouragement, and Mohammad Yousuf Jamal for his numerous informative contributions, collectively enriching the content of this paper. Additionally, I acknowledge the partial support from DAE-BRNS, India, Grant
No. 57/14/02/ 2021-BRNS. I also acknowledge partial support from the SERB Fellowship Project Code No.RD/0122-SERBF30-001.

\section{Appendix}
\label{AP}
Consider a partition $\{t_n = \frac{nT}{N},;0 \leq n \leq N\}$  of time interval $ [0,T]$. The three-step numerical scheme called L2 approximation is used for the case of superdiffusion. The second derivative $u^{(2)}$ is approximated using a central difference formula, and the resulting numerical scheme involves the values of $u$ at the previous three-time points $u^{n-1}$, $u^{n-2}$, and $u^{n-3}$.

\begin{widetext}
For $1<\nu\leq 2$, superdiffusion
\begin{align}
 \nonumber^{C} D^\nu_{0+}u(t_n)&= \frac{1}{\Gamma({2-\nu})}\int_0^{t_n} \frac{u^{(2)}(s)}{(t_n-s)^{\nu-1}}ds\\
\nonumber &= \frac{1}{\Gamma({2-\nu})}\sum_{j=1}^n\int_{t_{j-1}}^{t_j} \frac{u^{(2)}(s)}{(t_n-s)^{\nu-1}}ds\\
\nonumber &\approx \frac{1}{\Gamma({2-\nu})}\sum_{j=1}^n \frac{u(t_j)-2u(t_{j-1})+u(t_{j-2})}{\Delta t^2}\int_{t_{j-1}}^{t_j} \frac{1}{(t_n-s)^{\nu-1}}ds\\
\nonumber &= \frac{1}{\Gamma({3-\nu})}\sum_{j=1}^n\frac{((t_n - t_{j-1})^{2-\nu}- (t_n - t_{j})^{2-\nu})}{\Delta t^2} (u(t_j)-2u(t_{j-1})+u(t_{j-2})) \\
\nonumber &= \sum_{j=0}^{n-2}b_j (u(t_{n-j})-2u(t_{n-j-1})+u(t_{n-j-2}))\\
&+\frac{n^{2-\nu} - (n-1)^{2-\nu}}{\Gamma(3-\nu)\Delta t^{\nu}}(u(t_1) - u(t_0)).
 \end{align}


\begin{align}\label{SuperA}
 ^{C} D^\nu_{n}u(t_n)=\begin{cases}\displaystyle \frac{u^1-u^0}{\Delta t^{\nu}\Gamma(3-\nu)} \; &:\;n=1 \\
 \displaystyle\frac{u^2-2u^1 +u^0}{\Delta t^{\nu}\Gamma(3-\nu)} \; &:\;n=2 \\
 \displaystyle\frac{u^n-2u^{n-1}+u^{n-2}}{\Delta t^{\nu}\Gamma(3-\nu)} +\sum_{j=1}^{n-2} b_j(u^{n-j}-2u^{n-j-1} + u^{n-j-2}) & \;\\
 \displaystyle+\;\frac{n^{2-\nu} - (n-1)^{2-\nu}}{\Gamma(3-\nu)\Delta t^{\nu}}(u^1 - u^0)\; &:\;n \geq 3
 \end{cases}
\end{align}
\end{widetext}
where the coefficients $b_j= \frac{(j+1)^{2-\nu}-j^{2-\nu}}{\Gamma(3-\nu)\Delta t^{\nu}}$ in the scheme is determined by the difference formula for the second derivative and are used to account for the fractional order.

		



\bibliography{ref1}

\begin{thebibliography}{104}%
\makeatletter
\providecommand \@ifxundefined [1]{%
 \@ifx{#1\undefined}
}%
\providecommand \@ifnum [1]{%
 \ifnum #1\expandafter \@firstoftwo
 \else \expandafter \@secondoftwo
 \fi
}%
\providecommand \@ifx [1]{%
 \ifx #1\expandafter \@firstoftwo
 \else \expandafter \@secondoftwo
 \fi
}%
\providecommand \natexlab [1]{#1}%
\providecommand \enquote  [1]{``#1''}%
\providecommand \bibnamefont  [1]{#1}%
\providecommand \bibfnamefont [1]{#1}%
\providecommand \citenamefont [1]{#1}%
\providecommand \href@noop [0]{\@secondoftwo}%
\providecommand \href [0]{\begingroup \@sanitize@url \@href}%
\providecommand \@href[1]{\@@startlink{#1}\@@href}%
\providecommand \@@href[1]{\endgroup#1\@@endlink}%
\providecommand \@sanitize@url [0]{\catcode `\\12\catcode `\$12\catcode
  `\&12\catcode `\#12\catcode `\^12\catcode `\_12\catcode `\%12\relax}%
\providecommand \@@startlink[1]{}%
\providecommand \@@endlink[0]{}%
\providecommand \url  [0]{\begingroup\@sanitize@url \@url }%
\providecommand \@url [1]{\endgroup\@href {#1}{\urlprefix }}%
\providecommand \urlprefix  [0]{URL }%
\providecommand \Eprint [0]{\href }%
\providecommand \doibase [0]{http://dx.doi.org/}%
\providecommand \selectlanguage [0]{\@gobble}%
\providecommand \bibinfo  [0]{\@secondoftwo}%
\providecommand \bibfield  [0]{\@secondoftwo}%
\providecommand \translation [1]{[#1]}%
\providecommand \BibitemOpen [0]{}%
\providecommand \bibitemStop [0]{}%
\providecommand \bibitemNoStop [0]{.\EOS\space}%
\providecommand \EOS [0]{\spacefactor3000\relax}%
\providecommand \BibitemShut  [1]{\csname bibitem#1\endcsname}%
\let\auto@bib@innerbib\@empty
\bibitem [{\citenamefont {Adams}\ \emph {et~al.}(2006)\citenamefont {Adams}
  \emph {et~al.}}]{STAR:2006vcp}%
  \BibitemOpen
  \bibfield  {author} {\bibinfo {author} {\bibfnamefont {J.}~\bibnamefont
  {Adams}} \emph {et~al.} (\bibinfo {collaboration} {STAR}),\ }\href {\doibase
  10.1103/PhysRevLett.97.162301} {\bibfield  {journal} {\bibinfo  {journal}
  {Phys. Rev. Lett.}\ }\textbf {\bibinfo {volume} {97}},\ \bibinfo {pages}
  {162301} (\bibinfo {year} {2006})},\ \Eprint
  {http://arxiv.org/abs/nucl-ex/0604018} {arXiv:nucl-ex/0604018} \BibitemShut
  {NoStop}%
\bibitem [{\citenamefont {Adams}\ \emph {et~al.}(2005)\citenamefont {Adams}
  \emph {et~al.}}]{Adams:2005dq}%
  \BibitemOpen
  \bibfield  {author} {\bibinfo {author} {\bibfnamefont {J.}~\bibnamefont
  {Adams}} \emph {et~al.} (\bibinfo {collaboration} {STAR}),\ }\href {\doibase
  10.1016/j.nuclphysa.2005.03.085} {\bibfield  {journal} {\bibinfo  {journal}
  {Nucl. Phys. A}\ }\textbf {\bibinfo {volume} {757}},\ \bibinfo {pages} {102}
  (\bibinfo {year} {2005})},\ \Eprint {http://arxiv.org/abs/nucl-ex/0501009}
  {arXiv:nucl-ex/0501009} \BibitemShut {NoStop}%
\bibitem [{\citenamefont {Adcox}\ \emph {et~al.}(2005)\citenamefont {Adcox}
  \emph {et~al.}}]{PHENIX:2004vcz}%
  \BibitemOpen
  \bibfield  {author} {\bibinfo {author} {\bibfnamefont {K.}~\bibnamefont
  {Adcox}} \emph {et~al.} (\bibinfo {collaboration} {PHENIX}),\ }\href
  {\doibase 10.1016/j.nuclphysa.2005.03.086} {\bibfield  {journal} {\bibinfo
  {journal} {Nucl. Phys. A}\ }\textbf {\bibinfo {volume} {757}},\ \bibinfo
  {pages} {184} (\bibinfo {year} {2005})},\ \Eprint
  {http://arxiv.org/abs/nucl-ex/0410003} {arXiv:nucl-ex/0410003} \BibitemShut
  {NoStop}%
\bibitem [{\citenamefont {Aamodt}\ \emph {et~al.}(2010)\citenamefont {Aamodt}
  \emph {et~al.}}]{ALICE:2010khr}%
  \BibitemOpen
  \bibfield  {author} {\bibinfo {author} {\bibfnamefont {K.}~\bibnamefont
  {Aamodt}} \emph {et~al.} (\bibinfo {collaboration} {ALICE}),\ }\href
  {\doibase 10.1103/PhysRevLett.105.252301} {\bibfield  {journal} {\bibinfo
  {journal} {Phys. Rev. Lett.}\ }\textbf {\bibinfo {volume} {105}},\ \bibinfo
  {pages} {252301} (\bibinfo {year} {2010})},\ \Eprint
  {http://arxiv.org/abs/1011.3916} {arXiv:1011.3916 [nucl-ex]} \BibitemShut
  {NoStop}%
\bibitem [{\citenamefont {Arsene}\ \emph {et~al.}(2005)\citenamefont {Arsene}
  \emph {et~al.}}]{BRAHMS:2005gow}%
  \BibitemOpen
  \bibfield  {author} {\bibinfo {author} {\bibfnamefont {I.}~\bibnamefont
  {Arsene}} \emph {et~al.} (\bibinfo {collaboration} {BRAHMS}),\ }\href
  {\doibase 10.1103/PhysRevC.72.014908} {\bibfield  {journal} {\bibinfo
  {journal} {Phys. Rev. C}\ }\textbf {\bibinfo {volume} {72}},\ \bibinfo
  {pages} {014908} (\bibinfo {year} {2005})},\ \Eprint
  {http://arxiv.org/abs/nucl-ex/0503010} {arXiv:nucl-ex/0503010} \BibitemShut
  {NoStop}%
\bibitem [{\citenamefont {van Hees}\ and\ \citenamefont
  {Rapp}(2005)}]{vanHees:2004gq}%
  \BibitemOpen
  \bibfield  {author} {\bibinfo {author} {\bibfnamefont {H.}~\bibnamefont {van
  Hees}}\ and\ \bibinfo {author} {\bibfnamefont {R.}~\bibnamefont {Rapp}},\
  }\href {\doibase 10.1103/PhysRevC.71.034907} {\bibfield  {journal} {\bibinfo
  {journal} {Phys. Rev. C}\ }\textbf {\bibinfo {volume} {71}},\ \bibinfo
  {pages} {034907} (\bibinfo {year} {2005})},\ \Eprint
  {http://arxiv.org/abs/nucl-th/0412015} {arXiv:nucl-th/0412015} \BibitemShut
  {NoStop}%
\bibitem [{\citenamefont {Rapp}\ and\ \citenamefont {van
  Hees}(2010)}]{Rapp:2009my}%
  \BibitemOpen
  \bibfield  {author} {\bibinfo {author} {\bibfnamefont {R.}~\bibnamefont
  {Rapp}}\ and\ \bibinfo {author} {\bibfnamefont {H.}~\bibnamefont {van Hees}}\
  }(\bibinfo {year} {2010})\ pp.\ \bibinfo {pages} {111--206},\ \Eprint
  {http://arxiv.org/abs/0903.1096} {arXiv:0903.1096 [hep-ph]} \BibitemShut
  {NoStop}%
\bibitem [{\citenamefont {Song}\ \emph {et~al.}(2015)\citenamefont {Song},
  \citenamefont {Berrehrah}, \citenamefont {Cabrera}, \citenamefont
  {Torres-Rincon}, \citenamefont {Tolos}, \citenamefont {Cassing},\ and\
  \citenamefont {Bratkovskaya}}]{Song:2015sfa}%
  \BibitemOpen
  \bibfield  {author} {\bibinfo {author} {\bibfnamefont {T.}~\bibnamefont
  {Song}}, \bibinfo {author} {\bibfnamefont {H.}~\bibnamefont {Berrehrah}},
  \bibinfo {author} {\bibfnamefont {D.}~\bibnamefont {Cabrera}}, \bibinfo
  {author} {\bibfnamefont {J.~M.}\ \bibnamefont {Torres-Rincon}}, \bibinfo
  {author} {\bibfnamefont {L.}~\bibnamefont {Tolos}}, \bibinfo {author}
  {\bibfnamefont {W.}~\bibnamefont {Cassing}}, \ and\ \bibinfo {author}
  {\bibfnamefont {E.}~\bibnamefont {Bratkovskaya}},\ }\href {\doibase
  10.1103/PhysRevC.92.014910} {\bibfield  {journal} {\bibinfo  {journal} {Phys.
  Rev. C}\ }\textbf {\bibinfo {volume} {92}},\ \bibinfo {pages} {014910}
  (\bibinfo {year} {2015})},\ \Eprint {http://arxiv.org/abs/1503.03039}
  {arXiv:1503.03039 [nucl-th]} \BibitemShut {NoStop}%
\bibitem [{\citenamefont {Andronic}\ \emph {et~al.}(2016)\citenamefont
  {Andronic} \emph {et~al.}}]{Andronic:2015wma}%
  \BibitemOpen
  \bibfield  {author} {\bibinfo {author} {\bibfnamefont {A.}~\bibnamefont
  {Andronic}} \emph {et~al.},\ }\href {\doibase 10.1140/epjc/s10052-015-3819-5}
  {\bibfield  {journal} {\bibinfo  {journal} {Eur. Phys. J. C}\ }\textbf
  {\bibinfo {volume} {76}},\ \bibinfo {pages} {107} (\bibinfo {year} {2016})},\
  \Eprint {http://arxiv.org/abs/1506.03981} {arXiv:1506.03981 [nucl-ex]}
  \BibitemShut {NoStop}%
\bibitem [{\citenamefont {Dong}\ and\ \citenamefont
  {Greco}(2019)}]{Dong:2019unq}%
  \BibitemOpen
  \bibfield  {author} {\bibinfo {author} {\bibfnamefont {X.}~\bibnamefont
  {Dong}}\ and\ \bibinfo {author} {\bibfnamefont {V.}~\bibnamefont {Greco}},\
  }\href {\doibase 10.1016/j.ppnp.2018.08.001} {\bibfield  {journal} {\bibinfo
  {journal} {Prog. Part. Nucl. Phys.}\ }\textbf {\bibinfo {volume} {104}},\
  \bibinfo {pages} {97} (\bibinfo {year} {2019})}\BibitemShut {NoStop}%
\bibitem [{\citenamefont {Cao}\ \emph {et~al.}(2019)\citenamefont {Cao} \emph
  {et~al.}}]{Cao:2018ews}%
  \BibitemOpen
  \bibfield  {author} {\bibinfo {author} {\bibfnamefont {S.}~\bibnamefont
  {Cao}} \emph {et~al.},\ }\href {\doibase 10.1103/PhysRevC.99.054907}
  {\bibfield  {journal} {\bibinfo  {journal} {Phys. Rev. C}\ }\textbf {\bibinfo
  {volume} {99}},\ \bibinfo {pages} {054907} (\bibinfo {year} {2019})},\
  \Eprint {http://arxiv.org/abs/1809.07894} {arXiv:1809.07894 [nucl-th]}
  \BibitemShut {NoStop}%
\bibitem [{\citenamefont {Beraudo}\ \emph {et~al.}(2018)\citenamefont {Beraudo}
  \emph {et~al.}}]{Rapp:2018qla}%
  \BibitemOpen
  \bibfield  {author} {\bibinfo {author} {\bibfnamefont {A.}~\bibnamefont
  {Beraudo}} \emph {et~al.},\ }\href {\doibase 10.1016/j.nuclphysa.2018.09.002}
  {\bibfield  {journal} {\bibinfo  {journal} {Nucl. Phys. A}\ }\textbf
  {\bibinfo {volume} {979}},\ \bibinfo {pages} {21} (\bibinfo {year} {2018})},\
  \Eprint {http://arxiv.org/abs/1803.03824} {arXiv:1803.03824 [nucl-th]}
  \BibitemShut {NoStop}%
\bibitem [{\citenamefont {Prino}\ and\ \citenamefont
  {Rapp}(2016)}]{Prino:2016cni}%
  \BibitemOpen
  \bibfield  {author} {\bibinfo {author} {\bibfnamefont {F.}~\bibnamefont
  {Prino}}\ and\ \bibinfo {author} {\bibfnamefont {R.}~\bibnamefont {Rapp}},\
  }\href {\doibase 10.1088/0954-3899/43/9/093002} {\bibfield  {journal}
  {\bibinfo  {journal} {J. Phys. G}\ }\textbf {\bibinfo {volume} {43}},\
  \bibinfo {pages} {093002} (\bibinfo {year} {2016})},\ \Eprint
  {http://arxiv.org/abs/1603.00529} {arXiv:1603.00529 [nucl-ex]} \BibitemShut
  {NoStop}%
\bibitem [{\citenamefont {Aarts}\ \emph {et~al.}(2017)\citenamefont {Aarts}
  \emph {et~al.}}]{Aarts:2016hap}%
  \BibitemOpen
  \bibfield  {author} {\bibinfo {author} {\bibfnamefont {G.}~\bibnamefont
  {Aarts}} \emph {et~al.},\ }\href {\doibase 10.1140/epja/i2017-12282-9}
  {\bibfield  {journal} {\bibinfo  {journal} {Eur. Phys. J. A}\ }\textbf
  {\bibinfo {volume} {53}},\ \bibinfo {pages} {93} (\bibinfo {year} {2017})},\
  \Eprint {http://arxiv.org/abs/1612.08032} {arXiv:1612.08032 [nucl-th]}
  \BibitemShut {NoStop}%
\bibitem [{\citenamefont {Uphoff}\ \emph {et~al.}(2011)\citenamefont {Uphoff},
  \citenamefont {Fochler}, \citenamefont {Xu},\ and\ \citenamefont
  {Greiner}}]{Uphoff:2011ad}%
  \BibitemOpen
  \bibfield  {author} {\bibinfo {author} {\bibfnamefont {J.}~\bibnamefont
  {Uphoff}}, \bibinfo {author} {\bibfnamefont {O.}~\bibnamefont {Fochler}},
  \bibinfo {author} {\bibfnamefont {Z.}~\bibnamefont {Xu}}, \ and\ \bibinfo
  {author} {\bibfnamefont {C.}~\bibnamefont {Greiner}},\ }\href {\doibase
  10.1103/PhysRevC.84.024908} {\bibfield  {journal} {\bibinfo  {journal} {Phys.
  Rev. C}\ }\textbf {\bibinfo {volume} {84}},\ \bibinfo {pages} {024908}
  (\bibinfo {year} {2011})},\ \Eprint {http://arxiv.org/abs/1104.2295}
  {arXiv:1104.2295 [hep-ph]} \BibitemShut {NoStop}%
\bibitem [{\citenamefont {Golam~Mustafa}\ \emph {et~al.}(1998)\citenamefont
  {Golam~Mustafa}, \citenamefont {Pal},\ and\ \citenamefont
  {Kumar~Srivastava}}]{GolamMustafa:1997id}%
  \BibitemOpen
  \bibfield  {author} {\bibinfo {author} {\bibfnamefont {M.}~\bibnamefont
  {Golam~Mustafa}}, \bibinfo {author} {\bibfnamefont {D.}~\bibnamefont {Pal}},
  \ and\ \bibinfo {author} {\bibfnamefont {D.}~\bibnamefont
  {Kumar~Srivastava}},\ }\href {\doibase 10.1103/PhysRevC.57.3499} {\bibfield
  {journal} {\bibinfo  {journal} {Phys. Rev. C}\ }\textbf {\bibinfo {volume}
  {57}},\ \bibinfo {pages} {889} (\bibinfo {year} {1998})},\ \bibinfo {note}
  {[Erratum: Phys.Rev.C 57, 3499--3499 (1998)]},\ \Eprint
  {http://arxiv.org/abs/nucl-th/9706001} {arXiv:nucl-th/9706001} \BibitemShut
  {NoStop}%
\bibitem [{\citenamefont {Plumari}\ \emph {et~al.}(2018)\citenamefont
  {Plumari}, \citenamefont {Minissale}, \citenamefont {Das}, \citenamefont
  {Coci},\ and\ \citenamefont {Greco}}]{Plumari:2017ntm}%
  \BibitemOpen
  \bibfield  {author} {\bibinfo {author} {\bibfnamefont {S.}~\bibnamefont
  {Plumari}}, \bibinfo {author} {\bibfnamefont {V.}~\bibnamefont {Minissale}},
  \bibinfo {author} {\bibfnamefont {S.~K.}\ \bibnamefont {Das}}, \bibinfo
  {author} {\bibfnamefont {G.}~\bibnamefont {Coci}}, \ and\ \bibinfo {author}
  {\bibfnamefont {V.}~\bibnamefont {Greco}},\ }\href {\doibase
  10.1140/epjc/s10052-018-5828-7} {\bibfield  {journal} {\bibinfo  {journal}
  {Eur. Phys. J. C}\ }\textbf {\bibinfo {volume} {78}},\ \bibinfo {pages} {348}
  (\bibinfo {year} {2018})},\ \Eprint {http://arxiv.org/abs/1712.00730}
  {arXiv:1712.00730 [hep-ph]} \BibitemShut {NoStop}%
\bibitem [{\citenamefont {Gossiaux}\ and\ \citenamefont
  {Aichelin}(2008)}]{Gossiaux:2008jv}%
  \BibitemOpen
  \bibfield  {author} {\bibinfo {author} {\bibfnamefont {P.~B.}\ \bibnamefont
  {Gossiaux}}\ and\ \bibinfo {author} {\bibfnamefont {J.}~\bibnamefont
  {Aichelin}},\ }\href {\doibase 10.1103/PhysRevC.78.014904} {\bibfield
  {journal} {\bibinfo  {journal} {Phys. Rev. C}\ }\textbf {\bibinfo {volume}
  {78}},\ \bibinfo {pages} {014904} (\bibinfo {year} {2008})},\ \Eprint
  {http://arxiv.org/abs/0802.2525} {arXiv:0802.2525 [hep-ph]} \BibitemShut
  {NoStop}%
\bibitem [{\citenamefont {Prakash}\ \emph {et~al.}(2021)\citenamefont
  {Prakash}, \citenamefont {Kurian}, \citenamefont {Das},\ and\ \citenamefont
  {Chandra}}]{Prakash:2021lwt}%
  \BibitemOpen
  \bibfield  {author} {\bibinfo {author} {\bibfnamefont {J.}~\bibnamefont
  {Prakash}}, \bibinfo {author} {\bibfnamefont {M.}~\bibnamefont {Kurian}},
  \bibinfo {author} {\bibfnamefont {S.~K.}\ \bibnamefont {Das}}, \ and\
  \bibinfo {author} {\bibfnamefont {V.}~\bibnamefont {Chandra}},\ }\href
  {\doibase 10.1103/PhysRevD.103.094009} {\bibfield  {journal} {\bibinfo
  {journal} {Phys. Rev. D}\ }\textbf {\bibinfo {volume} {103}},\ \bibinfo
  {pages} {094009} (\bibinfo {year} {2021})},\ \Eprint
  {http://arxiv.org/abs/2102.07082} {arXiv:2102.07082 [hep-ph]} \BibitemShut
  {NoStop}%
\bibitem [{\citenamefont {Prakash}\ \emph {et~al.}(2023)\citenamefont
  {Prakash}, \citenamefont {Chandra},\ and\ \citenamefont
  {Das}}]{Prakash:2023wbs}%
  \BibitemOpen
  \bibfield  {author} {\bibinfo {author} {\bibfnamefont {J.}~\bibnamefont
  {Prakash}}, \bibinfo {author} {\bibfnamefont {V.}~\bibnamefont {Chandra}}, \
  and\ \bibinfo {author} {\bibfnamefont {S.~K.}\ \bibnamefont {Das}},\ }\href
  {\doibase 10.1103/PhysRevD.108.096016} {\bibfield  {journal} {\bibinfo
  {journal} {Phys. Rev. D}\ }\textbf {\bibinfo {volume} {108}},\ \bibinfo
  {pages} {096016} (\bibinfo {year} {2023})},\ \Eprint
  {http://arxiv.org/abs/2306.07966} {arXiv:2306.07966 [hep-ph]} \BibitemShut
  {NoStop}%
\bibitem [{\citenamefont {Prakash}\ and\ \citenamefont
  {Jamal}(2023{\natexlab{a}})}]{Prakash:2023hfj}%
  \BibitemOpen
  \bibfield  {author} {\bibinfo {author} {\bibfnamefont {J.}~\bibnamefont
  {Prakash}}\ and\ \bibinfo {author} {\bibfnamefont {M.~Y.}\ \bibnamefont
  {Jamal}},\ }\href@noop {} {\  (\bibinfo {year} {2023}{\natexlab{a}})},\
  \Eprint {http://arxiv.org/abs/2304.04003} {arXiv:2304.04003 [nucl-th]}
  \BibitemShut {NoStop}%
\bibitem [{\citenamefont {Singh}\ \emph {et~al.}(2023)\citenamefont {Singh},
  \citenamefont {Kurian}, \citenamefont {Jeon},\ and\ \citenamefont
  {Gale}}]{Singh:2023smw}%
  \BibitemOpen
  \bibfield  {author} {\bibinfo {author} {\bibfnamefont {M.}~\bibnamefont
  {Singh}}, \bibinfo {author} {\bibfnamefont {M.}~\bibnamefont {Kurian}},
  \bibinfo {author} {\bibfnamefont {S.}~\bibnamefont {Jeon}}, \ and\ \bibinfo
  {author} {\bibfnamefont {C.}~\bibnamefont {Gale}},\ }\href {\doibase
  10.1103/PhysRevC.108.054901} {\bibfield  {journal} {\bibinfo  {journal}
  {Phys. Rev. C}\ }\textbf {\bibinfo {volume} {108}},\ \bibinfo {pages}
  {054901} (\bibinfo {year} {2023})},\ \Eprint
  {http://arxiv.org/abs/2306.09514} {arXiv:2306.09514 [nucl-th]} \BibitemShut
  {NoStop}%
\bibitem [{\citenamefont {Kurian}\ \emph {et~al.}(2020)\citenamefont {Kurian},
  \citenamefont {Singh}, \citenamefont {Chandra}, \citenamefont {Jeon},\ and\
  \citenamefont {Gale}}]{Kurian:2020orp}%
  \BibitemOpen
  \bibfield  {author} {\bibinfo {author} {\bibfnamefont {M.}~\bibnamefont
  {Kurian}}, \bibinfo {author} {\bibfnamefont {M.}~\bibnamefont {Singh}},
  \bibinfo {author} {\bibfnamefont {V.}~\bibnamefont {Chandra}}, \bibinfo
  {author} {\bibfnamefont {S.}~\bibnamefont {Jeon}}, \ and\ \bibinfo {author}
  {\bibfnamefont {C.}~\bibnamefont {Gale}},\ }\href {\doibase
  10.1103/PhysRevC.102.044907} {\bibfield  {journal} {\bibinfo  {journal}
  {Phys. Rev. C}\ }\textbf {\bibinfo {volume} {102}},\ \bibinfo {pages}
  {044907} (\bibinfo {year} {2020})},\ \Eprint
  {http://arxiv.org/abs/2007.07705} {arXiv:2007.07705 [hep-ph]} \BibitemShut
  {NoStop}%
\bibitem [{\citenamefont {Cao}\ \emph {et~al.}(2016)\citenamefont {Cao},
  \citenamefont {Luo}, \citenamefont {Qin},\ and\ \citenamefont
  {Wang}}]{Cao:2016gvr}%
  \BibitemOpen
  \bibfield  {author} {\bibinfo {author} {\bibfnamefont {S.}~\bibnamefont
  {Cao}}, \bibinfo {author} {\bibfnamefont {T.}~\bibnamefont {Luo}}, \bibinfo
  {author} {\bibfnamefont {G.-Y.}\ \bibnamefont {Qin}}, \ and\ \bibinfo
  {author} {\bibfnamefont {X.-N.}\ \bibnamefont {Wang}},\ }\href {\doibase
  10.1103/PhysRevC.94.014909} {\bibfield  {journal} {\bibinfo  {journal} {Phys.
  Rev. C}\ }\textbf {\bibinfo {volume} {94}},\ \bibinfo {pages} {014909}
  (\bibinfo {year} {2016})},\ \Eprint {http://arxiv.org/abs/1605.06447}
  {arXiv:1605.06447 [nucl-th]} \BibitemShut {NoStop}%
\bibitem [{\citenamefont {Mazumder}\ \emph {et~al.}(2011)\citenamefont
  {Mazumder}, \citenamefont {Bhattacharyya}, \citenamefont {Alam},\ and\
  \citenamefont {Das}}]{Mazumder:2011nj}%
  \BibitemOpen
  \bibfield  {author} {\bibinfo {author} {\bibfnamefont {S.}~\bibnamefont
  {Mazumder}}, \bibinfo {author} {\bibfnamefont {T.}~\bibnamefont
  {Bhattacharyya}}, \bibinfo {author} {\bibfnamefont {J.-e.}\ \bibnamefont
  {Alam}}, \ and\ \bibinfo {author} {\bibfnamefont {S.~K.}\ \bibnamefont
  {Das}},\ }\href {\doibase 10.1103/PhysRevC.84.044901} {\bibfield  {journal}
  {\bibinfo  {journal} {Phys. Rev. C}\ }\textbf {\bibinfo {volume} {84}},\
  \bibinfo {pages} {044901} (\bibinfo {year} {2011})},\ \Eprint
  {http://arxiv.org/abs/1106.2615} {arXiv:1106.2615 [nucl-th]} \BibitemShut
  {NoStop}%
\bibitem [{\citenamefont {Zhang}\ \emph {et~al.}(2023)\citenamefont {Zhang},
  \citenamefont {Zheng}, \citenamefont {Shi},\ and\ \citenamefont
  {Lin}}]{Zhang:2022fum}%
  \BibitemOpen
  \bibfield  {author} {\bibinfo {author} {\bibfnamefont {C.}~\bibnamefont
  {Zhang}}, \bibinfo {author} {\bibfnamefont {L.}~\bibnamefont {Zheng}},
  \bibinfo {author} {\bibfnamefont {S.}~\bibnamefont {Shi}}, \ and\ \bibinfo
  {author} {\bibfnamefont {Z.-W.}\ \bibnamefont {Lin}},\ }\href {\doibase
  10.1016/j.physletb.2023.138219} {\bibfield  {journal} {\bibinfo  {journal}
  {Phys. Lett. B}\ }\textbf {\bibinfo {volume} {846}},\ \bibinfo {pages}
  {138219} (\bibinfo {year} {2023})},\ \Eprint
  {http://arxiv.org/abs/2210.07767} {arXiv:2210.07767 [nucl-th]} \BibitemShut
  {NoStop}%
\bibitem [{\citenamefont {Jamal}\ \emph {et~al.}(2021)\citenamefont {Jamal},
  \citenamefont {Das},\ and\ \citenamefont {Ruggieri}}]{PhysRevD.103.054030}%
  \BibitemOpen
  \bibfield  {author} {\bibinfo {author} {\bibfnamefont {M.~Y.}\ \bibnamefont
  {Jamal}}, \bibinfo {author} {\bibfnamefont {S.~K.}\ \bibnamefont {Das}}, \
  and\ \bibinfo {author} {\bibfnamefont {M.}~\bibnamefont {Ruggieri}},\ }\href
  {\doibase 10.1103/PhysRevD.103.054030} {\bibfield  {journal} {\bibinfo
  {journal} {Phys. Rev. D}\ }\textbf {\bibinfo {volume} {103}},\ \bibinfo
  {pages} {054030} (\bibinfo {year} {2021})}\BibitemShut {NoStop}%
\bibitem [{\citenamefont {Jamal}\ and\ \citenamefont
  {Mohanty}(2021{\natexlab{a}})}]{Jamal:2021btg}%
  \BibitemOpen
  \bibfield  {author} {\bibinfo {author} {\bibfnamefont {M.~Y.}\ \bibnamefont
  {Jamal}}\ and\ \bibinfo {author} {\bibfnamefont {B.}~\bibnamefont
  {Mohanty}},\ }\href {\doibase 10.1140/epjc/s10052-021-09418-9} {\bibfield
  {journal} {\bibinfo  {journal} {Eur. Phys. J. C}\ }\textbf {\bibinfo {volume}
  {81}},\ \bibinfo {pages} {616} (\bibinfo {year} {2021}{\natexlab{a}})},\
  \Eprint {http://arxiv.org/abs/2101.00164} {arXiv:2101.00164 [nucl-th]}
  \BibitemShut {NoStop}%
\bibitem [{\citenamefont {Jamal}\ and\ \citenamefont
  {Mohanty}(2021{\natexlab{b}})}]{Jamal:2020emj}%
  \BibitemOpen
  \bibfield  {author} {\bibinfo {author} {\bibfnamefont {M.~Y.}\ \bibnamefont
  {Jamal}}\ and\ \bibinfo {author} {\bibfnamefont {B.}~\bibnamefont
  {Mohanty}},\ }\href {\doibase 10.1140/epjp/s13360-021-01098-4} {\bibfield
  {journal} {\bibinfo  {journal} {Eur. Phys. J. Plus}\ }\textbf {\bibinfo
  {volume} {136}},\ \bibinfo {pages} {130} (\bibinfo {year}
  {2021}{\natexlab{b}})},\ \Eprint {http://arxiv.org/abs/2002.09230}
  {arXiv:2002.09230 [nucl-th]} \BibitemShut {NoStop}%
\bibitem [{\citenamefont {Sun}\ \emph {et~al.}(2023)\citenamefont {Sun},
  \citenamefont {Plumari},\ and\ \citenamefont {Das}}]{Sun:2023adv}%
  \BibitemOpen
  \bibfield  {author} {\bibinfo {author} {\bibfnamefont {Y.}~\bibnamefont
  {Sun}}, \bibinfo {author} {\bibfnamefont {S.}~\bibnamefont {Plumari}}, \ and\
  \bibinfo {author} {\bibfnamefont {S.~K.}\ \bibnamefont {Das}},\ }\href
  {\doibase 10.1016/j.physletb.2023.138043} {\bibfield  {journal} {\bibinfo
  {journal} {Phys. Lett. B}\ }\textbf {\bibinfo {volume} {843}},\ \bibinfo
  {pages} {138043} (\bibinfo {year} {2023})},\ \Eprint
  {http://arxiv.org/abs/2304.12792} {arXiv:2304.12792 [nucl-th]} \BibitemShut
  {NoStop}%
\bibitem [{\citenamefont {Plumari}\ \emph {et~al.}(2020)\citenamefont
  {Plumari}, \citenamefont {Coci}, \citenamefont {Minissale}, \citenamefont
  {Das}, \citenamefont {Sun},\ and\ \citenamefont {Greco}}]{Plumari:2019hzp}%
  \BibitemOpen
  \bibfield  {author} {\bibinfo {author} {\bibfnamefont {S.}~\bibnamefont
  {Plumari}}, \bibinfo {author} {\bibfnamefont {G.}~\bibnamefont {Coci}},
  \bibinfo {author} {\bibfnamefont {V.}~\bibnamefont {Minissale}}, \bibinfo
  {author} {\bibfnamefont {S.~K.}\ \bibnamefont {Das}}, \bibinfo {author}
  {\bibfnamefont {Y.}~\bibnamefont {Sun}}, \ and\ \bibinfo {author}
  {\bibfnamefont {V.}~\bibnamefont {Greco}},\ }\href {\doibase
  10.1016/j.physletb.2020.135460} {\bibfield  {journal} {\bibinfo  {journal}
  {Phys. Lett. B}\ }\textbf {\bibinfo {volume} {805}},\ \bibinfo {pages}
  {135460} (\bibinfo {year} {2020})},\ \Eprint
  {http://arxiv.org/abs/1912.09350} {arXiv:1912.09350 [hep-ph]} \BibitemShut
  {NoStop}%
\bibitem [{\citenamefont {Prakash}\ and\ \citenamefont
  {Jamal}(2023{\natexlab{b}})}]{Prakash_2024}%
  \BibitemOpen
  \bibfield  {author} {\bibinfo {author} {\bibfnamefont {J.}~\bibnamefont
  {Prakash}}\ and\ \bibinfo {author} {\bibfnamefont {M.~Y.}\ \bibnamefont
  {Jamal}},\ }\href {\doibase 10.1088/1361-6471/ad10c9} {\bibfield  {journal}
  {\bibinfo  {journal} {Journal of Physics G: Nuclear and Particle Physics}\
  }\textbf {\bibinfo {volume} {51}},\ \bibinfo {pages} {025101} (\bibinfo
  {year} {2023}{\natexlab{b}})}\BibitemShut {NoStop}%
\bibitem [{\citenamefont {Debnath}\ \emph {et~al.}(2024)\citenamefont
  {Debnath}, \citenamefont {Ghosh}, \citenamefont {Jamal}, \citenamefont
  {Kurian},\ and\ \citenamefont {Prakash}}]{Debnath:2023zet}%
  \BibitemOpen
  \bibfield  {author} {\bibinfo {author} {\bibfnamefont {M.}~\bibnamefont
  {Debnath}}, \bibinfo {author} {\bibfnamefont {R.}~\bibnamefont {Ghosh}},
  \bibinfo {author} {\bibfnamefont {M.~Y.}\ \bibnamefont {Jamal}}, \bibinfo
  {author} {\bibfnamefont {M.}~\bibnamefont {Kurian}}, \ and\ \bibinfo {author}
  {\bibfnamefont {J.}~\bibnamefont {Prakash}},\ }\href {\doibase
  10.1103/PhysRevD.109.L011503} {\bibfield  {journal} {\bibinfo  {journal}
  {Phys. Rev. D}\ }\textbf {\bibinfo {volume} {109}},\ \bibinfo {pages}
  {L011503} (\bibinfo {year} {2024})},\ \Eprint
  {http://arxiv.org/abs/2311.16005} {arXiv:2311.16005 [hep-ph]} \BibitemShut
  {NoStop}%
\bibitem [{\citenamefont {Du}\ and\ \citenamefont
  {Qian}(2023)}]{du2023accelerated}%
  \BibitemOpen
  \bibfield  {author} {\bibinfo {author} {\bibfnamefont {X.}~\bibnamefont
  {Du}}\ and\ \bibinfo {author} {\bibfnamefont {W.}~\bibnamefont {Qian}},\
  }\href@noop {} {\bibfield  {journal} {\bibinfo  {journal} {arXiv preprint
  arXiv:2312.16294}\ } (\bibinfo {year} {2023})}\BibitemShut {NoStop}%
\bibitem [{\citenamefont {Shaikh}\ \emph {et~al.}(2021)\citenamefont {Shaikh},
  \citenamefont {Kurian}, \citenamefont {Das}, \citenamefont {Chandra},
  \citenamefont {Dash},\ and\ \citenamefont {Nandi}}]{Shaikh:2021lka}%
  \BibitemOpen
  \bibfield  {author} {\bibinfo {author} {\bibfnamefont {A.}~\bibnamefont
  {Shaikh}}, \bibinfo {author} {\bibfnamefont {M.}~\bibnamefont {Kurian}},
  \bibinfo {author} {\bibfnamefont {S.~K.}\ \bibnamefont {Das}}, \bibinfo
  {author} {\bibfnamefont {V.}~\bibnamefont {Chandra}}, \bibinfo {author}
  {\bibfnamefont {S.}~\bibnamefont {Dash}}, \ and\ \bibinfo {author}
  {\bibfnamefont {B.~K.}\ \bibnamefont {Nandi}},\ }\href {\doibase
  10.1103/PhysRevD.104.034017} {\bibfield  {journal} {\bibinfo  {journal}
  {Phys. Rev. D}\ }\textbf {\bibinfo {volume} {104}},\ \bibinfo {pages}
  {034017} (\bibinfo {year} {2021})},\ \Eprint
  {http://arxiv.org/abs/2105.14296} {arXiv:2105.14296 [hep-ph]} \BibitemShut
  {NoStop}%
\bibitem [{\citenamefont {Kumar}\ \emph {et~al.}(2022)\citenamefont {Kumar},
  \citenamefont {Kurian}, \citenamefont {Das},\ and\ \citenamefont
  {Chandra}}]{Kumar:2021goi}%
  \BibitemOpen
  \bibfield  {author} {\bibinfo {author} {\bibfnamefont {A.}~\bibnamefont
  {Kumar}}, \bibinfo {author} {\bibfnamefont {M.}~\bibnamefont {Kurian}},
  \bibinfo {author} {\bibfnamefont {S.~K.}\ \bibnamefont {Das}}, \ and\
  \bibinfo {author} {\bibfnamefont {V.}~\bibnamefont {Chandra}},\ }\href
  {\doibase 10.1103/PhysRevC.105.054903} {\bibfield  {journal} {\bibinfo
  {journal} {Phys. Rev. C}\ }\textbf {\bibinfo {volume} {105}},\ \bibinfo
  {pages} {054903} (\bibinfo {year} {2022})},\ \Eprint
  {http://arxiv.org/abs/2111.07563} {arXiv:2111.07563 [hep-ph]} \BibitemShut
  {NoStop}%
\bibitem [{\citenamefont {Das}\ \emph {et~al.}(2022)\citenamefont {Das} \emph
  {et~al.}}]{das2022dynamics}%
  \BibitemOpen
  \bibfield  {author} {\bibinfo {author} {\bibfnamefont {S.~K.}\ \bibnamefont
  {Das}} \emph {et~al.},\ }\href@noop {} {\bibfield  {journal} {\bibinfo
  {journal} {International Journal of Modern Physics E}\ ,\ \bibinfo {pages}
  {2250097}} (\bibinfo {year} {2022})}\BibitemShut {NoStop}%
\bibitem [{\citenamefont {Kubo}\ \emph {et~al.}(2012)\citenamefont {Kubo},
  \citenamefont {Toda},\ and\ \citenamefont
  {Hashitsume}}]{kubo2012statistical}%
  \BibitemOpen
  \bibfield  {author} {\bibinfo {author} {\bibfnamefont {R.}~\bibnamefont
  {Kubo}}, \bibinfo {author} {\bibfnamefont {M.}~\bibnamefont {Toda}}, \ and\
  \bibinfo {author} {\bibfnamefont {N.}~\bibnamefont {Hashitsume}},\
  }\href@noop {} {\ \textbf {\bibinfo {volume} {31}} (\bibinfo {year}
  {2012})}\BibitemShut {NoStop}%
\bibitem [{\citenamefont {Young}\ \emph {et~al.}(2012)\citenamefont {Young},
  \citenamefont {Schenke}, \citenamefont {Jeon},\ and\ \citenamefont
  {Gale}}]{PhysRevC.86.034905}%
  \BibitemOpen
  \bibfield  {author} {\bibinfo {author} {\bibfnamefont {C.}~\bibnamefont
  {Young}}, \bibinfo {author} {\bibfnamefont {B.}~\bibnamefont {Schenke}},
  \bibinfo {author} {\bibfnamefont {S.}~\bibnamefont {Jeon}}, \ and\ \bibinfo
  {author} {\bibfnamefont {C.}~\bibnamefont {Gale}},\ }\href {\doibase
  10.1103/PhysRevC.86.034905} {\bibfield  {journal} {\bibinfo  {journal} {Phys.
  Rev. C}\ }\textbf {\bibinfo {volume} {86}},\ \bibinfo {pages} {034905}
  (\bibinfo {year} {2012})}\BibitemShut {NoStop}%
\bibitem [{\citenamefont {Lang}\ \emph {et~al.}(2016)\citenamefont {Lang},
  \citenamefont {van Hees}, \citenamefont {Inghirami}, \citenamefont
  {Steinheimer},\ and\ \citenamefont {Bleicher}}]{PhysRevC.93.014901}%
  \BibitemOpen
  \bibfield  {author} {\bibinfo {author} {\bibfnamefont {T.}~\bibnamefont
  {Lang}}, \bibinfo {author} {\bibfnamefont {H.}~\bibnamefont {van Hees}},
  \bibinfo {author} {\bibfnamefont {G.}~\bibnamefont {Inghirami}}, \bibinfo
  {author} {\bibfnamefont {J.}~\bibnamefont {Steinheimer}}, \ and\ \bibinfo
  {author} {\bibfnamefont {M.}~\bibnamefont {Bleicher}},\ }\href {\doibase
  10.1103/PhysRevC.93.014901} {\bibfield  {journal} {\bibinfo  {journal} {Phys.
  Rev. C}\ }\textbf {\bibinfo {volume} {93}},\ \bibinfo {pages} {014901}
  (\bibinfo {year} {2016})}\BibitemShut {NoStop}%
\bibitem [{\citenamefont {Cao}\ and\ \citenamefont
  {Bass}(2011)}]{PhysRevC.84.064902}%
  \BibitemOpen
  \bibfield  {author} {\bibinfo {author} {\bibfnamefont {S.}~\bibnamefont
  {Cao}}\ and\ \bibinfo {author} {\bibfnamefont {S.~A.}\ \bibnamefont {Bass}},\
  }\href {\doibase 10.1103/PhysRevC.84.064902} {\bibfield  {journal} {\bibinfo
  {journal} {Phys. Rev. C}\ }\textbf {\bibinfo {volume} {84}},\ \bibinfo
  {pages} {064902} (\bibinfo {year} {2011})}\BibitemShut {NoStop}%
\bibitem [{\citenamefont {van Hees}\ \emph {et~al.}(2008)\citenamefont {van
  Hees}, \citenamefont {Mannarelli}, \citenamefont {Greco},\ and\ \citenamefont
  {Rapp}}]{vanHees:2007me}%
  \BibitemOpen
  \bibfield  {author} {\bibinfo {author} {\bibfnamefont {H.}~\bibnamefont {van
  Hees}}, \bibinfo {author} {\bibfnamefont {M.}~\bibnamefont {Mannarelli}},
  \bibinfo {author} {\bibfnamefont {V.}~\bibnamefont {Greco}}, \ and\ \bibinfo
  {author} {\bibfnamefont {R.}~\bibnamefont {Rapp}},\ }\href {\doibase
  10.1103/PhysRevLett.100.192301} {\bibfield  {journal} {\bibinfo  {journal}
  {Phys. Rev. Lett.}\ }\textbf {\bibinfo {volume} {100}},\ \bibinfo {pages}
  {192301} (\bibinfo {year} {2008})},\ \Eprint {http://arxiv.org/abs/0709.2884}
  {arXiv:0709.2884 [hep-ph]} \BibitemShut {NoStop}%
\bibitem [{\citenamefont {Das}\ \emph {et~al.}(2014)\citenamefont {Das},
  \citenamefont {Scardina}, \citenamefont {Plumari},\ and\ \citenamefont
  {Greco}}]{Das:2013kea}%
  \BibitemOpen
  \bibfield  {author} {\bibinfo {author} {\bibfnamefont {S.~K.}\ \bibnamefont
  {Das}}, \bibinfo {author} {\bibfnamefont {F.}~\bibnamefont {Scardina}},
  \bibinfo {author} {\bibfnamefont {S.}~\bibnamefont {Plumari}}, \ and\
  \bibinfo {author} {\bibfnamefont {V.}~\bibnamefont {Greco}},\ }\href
  {\doibase 10.1103/PhysRevC.90.044901} {\bibfield  {journal} {\bibinfo
  {journal} {Phys. Rev. C}\ }\textbf {\bibinfo {volume} {90}},\ \bibinfo
  {pages} {044901} (\bibinfo {year} {2014})},\ \Eprint
  {http://arxiv.org/abs/1312.6857} {arXiv:1312.6857 [nucl-th]} \BibitemShut
  {NoStop}%
\bibitem [{\citenamefont {He}\ \emph {et~al.}(2023)\citenamefont {He},
  \citenamefont {van Hees},\ and\ \citenamefont {Rapp}}]{He:2022ywp}%
  \BibitemOpen
  \bibfield  {author} {\bibinfo {author} {\bibfnamefont {M.}~\bibnamefont
  {He}}, \bibinfo {author} {\bibfnamefont {H.}~\bibnamefont {van Hees}}, \ and\
  \bibinfo {author} {\bibfnamefont {R.}~\bibnamefont {Rapp}},\ }\href {\doibase
  10.1016/j.ppnp.2023.104020} {\bibfield  {journal} {\bibinfo  {journal} {Prog.
  Part. Nucl. Phys.}\ }\textbf {\bibinfo {volume} {130}},\ \bibinfo {pages}
  {104020} (\bibinfo {year} {2023})},\ \Eprint
  {http://arxiv.org/abs/2204.09299} {arXiv:2204.09299 [hep-ph]} \BibitemShut
  {NoStop}%
\bibitem [{\citenamefont {He}\ \emph {et~al.}(2012)\citenamefont {He},
  \citenamefont {Fries},\ and\ \citenamefont {Rapp}}]{PhysRevC.86.014903}%
  \BibitemOpen
  \bibfield  {author} {\bibinfo {author} {\bibfnamefont {M.}~\bibnamefont
  {He}}, \bibinfo {author} {\bibfnamefont {R.~J.}\ \bibnamefont {Fries}}, \
  and\ \bibinfo {author} {\bibfnamefont {R.}~\bibnamefont {Rapp}},\ }\href
  {\doibase 10.1103/PhysRevC.86.014903} {\bibfield  {journal} {\bibinfo
  {journal} {Phys. Rev. C}\ }\textbf {\bibinfo {volume} {86}},\ \bibinfo
  {pages} {014903} (\bibinfo {year} {2012})}\BibitemShut {NoStop}%
\bibitem [{\citenamefont {Das}\ \emph {et~al.}(2015)\citenamefont {Das},
  \citenamefont {Scardina}, \citenamefont {Plumari},\ and\ \citenamefont
  {Greco}}]{Das:2015ana}%
  \BibitemOpen
  \bibfield  {author} {\bibinfo {author} {\bibfnamefont {S.~K.}\ \bibnamefont
  {Das}}, \bibinfo {author} {\bibfnamefont {F.}~\bibnamefont {Scardina}},
  \bibinfo {author} {\bibfnamefont {S.}~\bibnamefont {Plumari}}, \ and\
  \bibinfo {author} {\bibfnamefont {V.}~\bibnamefont {Greco}},\ }\href
  {\doibase 10.1016/j.physletb.2015.06.003} {\bibfield  {journal} {\bibinfo
  {journal} {Phys. Lett. B}\ }\textbf {\bibinfo {volume} {747}},\ \bibinfo
  {pages} {260} (\bibinfo {year} {2015})},\ \Eprint
  {http://arxiv.org/abs/1502.03757} {arXiv:1502.03757 [nucl-th]} \BibitemShut
  {NoStop}%
\bibitem [{\citenamefont {Balescu}(1995)}]{balescu1995anomalous}%
  \BibitemOpen
  \bibfield  {author} {\bibinfo {author} {\bibfnamefont {R.}~\bibnamefont
  {Balescu}},\ }\href {\doibase 10.1103/PhysRevE.51.4807} {\bibfield  {journal}
  {\bibinfo  {journal} {Physical Review E}\ }\textbf {\bibinfo {volume} {51}},\
  \bibinfo {pages} {4807} (\bibinfo {year} {1995})}\BibitemShut {NoStop}%
\bibitem [{\citenamefont {Bouchaud}\ and\ \citenamefont
  {Georges}(1990)}]{bouchaud1990anomalous}%
  \BibitemOpen
  \bibfield  {author} {\bibinfo {author} {\bibfnamefont {J.-P.}\ \bibnamefont
  {Bouchaud}}\ and\ \bibinfo {author} {\bibfnamefont {A.}~\bibnamefont
  {Georges}},\ }\href@noop {} {\bibfield  {journal} {\bibinfo  {journal}
  {Physics reports}\ }\textbf {\bibinfo {volume} {195}},\ \bibinfo {pages}
  {127} (\bibinfo {year} {1990})}\BibitemShut {NoStop}%
\bibitem [{\citenamefont {Metzler}\ and\ \citenamefont
  {Klafter}(2000)}]{metzler2000random}%
  \BibitemOpen
  \bibfield  {author} {\bibinfo {author} {\bibfnamefont {R.}~\bibnamefont
  {Metzler}}\ and\ \bibinfo {author} {\bibfnamefont {J.}~\bibnamefont
  {Klafter}},\ }\href@noop {} {\bibfield  {journal} {\bibinfo  {journal}
  {Physics reports}\ }\textbf {\bibinfo {volume} {339}},\ \bibinfo {pages} {1}
  (\bibinfo {year} {2000})}\BibitemShut {NoStop}%
\bibitem [{\citenamefont {Caspi}\ \emph {et~al.}(2000)\citenamefont {Caspi},
  \citenamefont {Granek},\ and\ \citenamefont {Elbaum}}]{PhysRevLett.85.5655}%
  \BibitemOpen
  \bibfield  {author} {\bibinfo {author} {\bibfnamefont {A.}~\bibnamefont
  {Caspi}}, \bibinfo {author} {\bibfnamefont {R.}~\bibnamefont {Granek}}, \
  and\ \bibinfo {author} {\bibfnamefont {M.}~\bibnamefont {Elbaum}},\ }\href
  {\doibase 10.1103/PhysRevLett.85.5655} {\bibfield  {journal} {\bibinfo
  {journal} {Phys. Rev. Lett.}\ }\textbf {\bibinfo {volume} {85}},\ \bibinfo
  {pages} {5655} (\bibinfo {year} {2000})}\BibitemShut {NoStop}%
\bibitem [{\citenamefont {Weber}\ \emph {et~al.}(2010)\citenamefont {Weber},
  \citenamefont {Spakowitz},\ and\ \citenamefont
  {Theriot}}]{PhysRevLett.104.238102}%
  \BibitemOpen
  \bibfield  {author} {\bibinfo {author} {\bibfnamefont {S.~C.}\ \bibnamefont
  {Weber}}, \bibinfo {author} {\bibfnamefont {A.~J.}\ \bibnamefont
  {Spakowitz}}, \ and\ \bibinfo {author} {\bibfnamefont {J.~A.}\ \bibnamefont
  {Theriot}},\ }\href {\doibase 10.1103/PhysRevLett.104.238102} {\bibfield
  {journal} {\bibinfo  {journal} {Phys. Rev. Lett.}\ }\textbf {\bibinfo
  {volume} {104}},\ \bibinfo {pages} {238102} (\bibinfo {year}
  {2010})}\BibitemShut {NoStop}%
\bibitem [{\citenamefont {Vitali}\ \emph {et~al.}(2018)\citenamefont {Vitali},
  \citenamefont {Sposini}, \citenamefont {Sliusarenko}, \citenamefont
  {Paradisi}, \citenamefont {Castellani},\ and\ \citenamefont
  {Pagnini}}]{vitali2018langevin}%
  \BibitemOpen
  \bibfield  {author} {\bibinfo {author} {\bibfnamefont {S.}~\bibnamefont
  {Vitali}}, \bibinfo {author} {\bibfnamefont {V.}~\bibnamefont {Sposini}},
  \bibinfo {author} {\bibfnamefont {O.}~\bibnamefont {Sliusarenko}}, \bibinfo
  {author} {\bibfnamefont {P.}~\bibnamefont {Paradisi}}, \bibinfo {author}
  {\bibfnamefont {G.}~\bibnamefont {Castellani}}, \ and\ \bibinfo {author}
  {\bibfnamefont {G.}~\bibnamefont {Pagnini}},\ }\href@noop {} {\bibfield
  {journal} {\bibinfo  {journal} {Journal of The Royal Society Interface}\
  }\textbf {\bibinfo {volume} {15}},\ \bibinfo {pages} {20180282} (\bibinfo
  {year} {2018})}\BibitemShut {NoStop}%
\bibitem [{\citenamefont {Oliveira}\ \emph {et~al.}(2019)\citenamefont
  {Oliveira}, \citenamefont {Ferreira}, \citenamefont {Lapas},\ and\
  \citenamefont {Vainstein}}]{oliveira2019anomalous}%
  \BibitemOpen
  \bibfield  {author} {\bibinfo {author} {\bibfnamefont {F.~A.}\ \bibnamefont
  {Oliveira}}, \bibinfo {author} {\bibfnamefont {R.~M.}\ \bibnamefont
  {Ferreira}}, \bibinfo {author} {\bibfnamefont {L.~C.}\ \bibnamefont {Lapas}},
  \ and\ \bibinfo {author} {\bibfnamefont {M.~H.}\ \bibnamefont {Vainstein}},\
  }\href@noop {} {\bibfield  {journal} {\bibinfo  {journal} {Frontiers in
  Physics}\ }\textbf {\bibinfo {volume} {7}},\ \bibinfo {pages} {18} (\bibinfo
  {year} {2019})}\BibitemShut {NoStop}%
\bibitem [{\citenamefont {Sokolov}\ and\ \citenamefont
  {Klafter}(2005)}]{sokolov2005diffusion}%
  \BibitemOpen
  \bibfield  {author} {\bibinfo {author} {\bibfnamefont {I.~M.}\ \bibnamefont
  {Sokolov}}\ and\ \bibinfo {author} {\bibfnamefont {J.}~\bibnamefont
  {Klafter}},\ }\href@noop {} {\bibfield  {journal} {\bibinfo  {journal}
  {Chaos: An Interdisciplinary Journal of Nonlinear Science}\ }\textbf
  {\bibinfo {volume} {15}} (\bibinfo {year} {2005})}\BibitemShut {NoStop}%
\bibitem [{\citenamefont {Lutz}(2001)}]{lutz2001fractional}%
  \BibitemOpen
  \bibfield  {author} {\bibinfo {author} {\bibfnamefont {E.}~\bibnamefont
  {Lutz}},\ }\href {\doibase 10.1103/PhysRevE.64.051106} {\bibfield  {journal}
  {\bibinfo  {journal} {Phys. Rev. E}\ }\textbf {\bibinfo {volume} {64}},\
  \bibinfo {pages} {051106} (\bibinfo {year} {2001})}\BibitemShut {NoStop}%
\bibitem [{\citenamefont {Lim}\ and\ \citenamefont
  {Teo}(2009)}]{lim2009modeling}%
  \BibitemOpen
  \bibfield  {author} {\bibinfo {author} {\bibfnamefont {S.}~\bibnamefont
  {Lim}}\ and\ \bibinfo {author} {\bibfnamefont {L.}~\bibnamefont {Teo}},\
  }\href@noop {} {\bibfield  {journal} {\bibinfo  {journal} {Journal of
  Statistical Mechanics: Theory and Experiment}\ }\textbf {\bibinfo {volume}
  {2009}},\ \bibinfo {pages} {P08015} (\bibinfo {year} {2009})}\BibitemShut
  {NoStop}%
\bibitem [{\citenamefont {Wang}\ and\ \citenamefont
  {Chen}(2023)}]{PhysRevE.107.024105}%
  \BibitemOpen
  \bibfield  {author} {\bibinfo {author} {\bibfnamefont {X.}~\bibnamefont
  {Wang}}\ and\ \bibinfo {author} {\bibfnamefont {Y.}~\bibnamefont {Chen}},\
  }\href {\doibase 10.1103/PhysRevE.107.024105} {\bibfield  {journal} {\bibinfo
   {journal} {Phys. Rev. E}\ }\textbf {\bibinfo {volume} {107}},\ \bibinfo
  {pages} {024105} (\bibinfo {year} {2023})}\BibitemShut {NoStop}%
\bibitem [{\citenamefont {Liang}\ \emph {et~al.}(2023)\citenamefont {Liang},
  \citenamefont {Wang}, \citenamefont {Metzler},\ and\ \citenamefont
  {Cherstvy}}]{PhysRevE.108.034113}%
  \BibitemOpen
  \bibfield  {author} {\bibinfo {author} {\bibfnamefont {Y.}~\bibnamefont
  {Liang}}, \bibinfo {author} {\bibfnamefont {W.}~\bibnamefont {Wang}},
  \bibinfo {author} {\bibfnamefont {R.}~\bibnamefont {Metzler}}, \ and\
  \bibinfo {author} {\bibfnamefont {A.~G.}\ \bibnamefont {Cherstvy}},\ }\href
  {\doibase 10.1103/PhysRevE.108.034113} {\bibfield  {journal} {\bibinfo
  {journal} {Phys. Rev. E}\ }\textbf {\bibinfo {volume} {108}},\ \bibinfo
  {pages} {034113} (\bibinfo {year} {2023})}\BibitemShut {NoStop}%
\bibitem [{\citenamefont {Wang}\ and\ \citenamefont
  {Barkai}(2020)}]{PhysRevLett.125.240606}%
  \BibitemOpen
  \bibfield  {author} {\bibinfo {author} {\bibfnamefont {W.}~\bibnamefont
  {Wang}}\ and\ \bibinfo {author} {\bibfnamefont {E.}~\bibnamefont {Barkai}},\
  }\href {\doibase 10.1103/PhysRevLett.125.240606} {\bibfield  {journal}
  {\bibinfo  {journal} {Phys. Rev. Lett.}\ }\textbf {\bibinfo {volume} {125}},\
  \bibinfo {pages} {240606} (\bibinfo {year} {2020})}\BibitemShut {NoStop}%
\bibitem [{\citenamefont {Cherstvy}\ \emph {et~al.}(2021)\citenamefont
  {Cherstvy}, \citenamefont {Wang}, \citenamefont {Metzler},\ and\
  \citenamefont {Sokolov}}]{PhysRevE.104.024115}%
  \BibitemOpen
  \bibfield  {author} {\bibinfo {author} {\bibfnamefont {A.~G.}\ \bibnamefont
  {Cherstvy}}, \bibinfo {author} {\bibfnamefont {W.}~\bibnamefont {Wang}},
  \bibinfo {author} {\bibfnamefont {R.}~\bibnamefont {Metzler}}, \ and\
  \bibinfo {author} {\bibfnamefont {I.~M.}\ \bibnamefont {Sokolov}},\ }\href
  {\doibase 10.1103/PhysRevE.104.024115} {\bibfield  {journal} {\bibinfo
  {journal} {Phys. Rev. E}\ }\textbf {\bibinfo {volume} {104}},\ \bibinfo
  {pages} {024115} (\bibinfo {year} {2021})}\BibitemShut {NoStop}%
\bibitem [{\citenamefont {Li}\ \emph {et~al.}(2012)\citenamefont {Li},
  \citenamefont {Zeng},\ and\ \citenamefont {Liu}}]{li2012spectral}%
  \BibitemOpen
  \bibfield  {author} {\bibinfo {author} {\bibfnamefont {C.}~\bibnamefont
  {Li}}, \bibinfo {author} {\bibfnamefont {F.}~\bibnamefont {Zeng}}, \ and\
  \bibinfo {author} {\bibfnamefont {F.}~\bibnamefont {Liu}},\ }\href@noop {}
  {\bibfield  {journal} {\bibinfo  {journal} {Fractional Calculus and Applied
  Analysis}\ }\textbf {\bibinfo {volume} {15}},\ \bibinfo {pages} {383}
  (\bibinfo {year} {2012})}\BibitemShut {NoStop}%
\bibitem [{\citenamefont {Kobelev}\ and\ \citenamefont
  {Romanov}(2000)}]{kobelev2000fractional}%
  \BibitemOpen
  \bibfield  {author} {\bibinfo {author} {\bibfnamefont {V.}~\bibnamefont
  {Kobelev}}\ and\ \bibinfo {author} {\bibfnamefont {E.}~\bibnamefont
  {Romanov}},\ }\href@noop {} {\bibfield  {journal} {\bibinfo  {journal}
  {Progress of Theoretical Physics Supplement}\ }\textbf {\bibinfo {volume}
  {139}},\ \bibinfo {pages} {470} (\bibinfo {year} {2000})}\BibitemShut
  {NoStop}%
\bibitem [{\citenamefont {Coffey}\ \emph {et~al.}(2004)\citenamefont {Coffey},
  \citenamefont {Kalmykov},\ and\ \citenamefont
  {Waldron}}]{coffey2012langevin}%
  \BibitemOpen
  \bibfield  {author} {\bibinfo {author} {\bibfnamefont {W.~T.}\ \bibnamefont
  {Coffey}}, \bibinfo {author} {\bibfnamefont {Y.~P.}\ \bibnamefont
  {Kalmykov}}, \ and\ \bibinfo {author} {\bibfnamefont {J.~T.}\ \bibnamefont
  {Waldron}},\ }\href {\doibase 10.1142/5343} {\bibfield  {journal} {\bibinfo
  {journal} {World Scientific Publishing Co., Inc., River Edge, NJ}\ }\textbf
  {\bibinfo {volume} {105}} (\bibinfo {year} {2004}),\
  10.1142/5343}\BibitemShut {NoStop}%
\bibitem [{\citenamefont {Megías}\ \emph {et~al.}(2024)\citenamefont
  {Megías}, \citenamefont {{Khalili Golmankhaneh}},\ and\ \citenamefont
  {Deppman}}]{MEGIAS2024138370}%
  \BibitemOpen
  \bibfield  {author} {\bibinfo {author} {\bibfnamefont {E.}~\bibnamefont
  {Megías}}, \bibinfo {author} {\bibfnamefont {A.}~\bibnamefont {{Khalili
  Golmankhaneh}}}, \ and\ \bibinfo {author} {\bibfnamefont {A.}~\bibnamefont
  {Deppman}},\ }\href {\doibase https://doi.org/10.1016/j.physletb.2023.138370}
  {\bibfield  {journal} {\bibinfo  {journal} {Physics Letters B}\ }\textbf
  {\bibinfo {volume} {848}},\ \bibinfo {pages} {138370} (\bibinfo {year}
  {2024})}\BibitemShut {NoStop}%
\bibitem [{\citenamefont {Gorenflo}\ and\ \citenamefont
  {Mainardi}(1997)}]{MR1611585}%
  \BibitemOpen
  \bibfield  {author} {\bibinfo {author} {\bibfnamefont {R.}~\bibnamefont
  {Gorenflo}}\ and\ \bibinfo {author} {\bibfnamefont {F.}~\bibnamefont
  {Mainardi}},\ }\href@noop {} {\ \bibinfo {series} {CISM Courses and Lect.},\
  \textbf {\bibinfo {volume} {378}},\ \bibinfo {pages} {223} (\bibinfo {year}
  {1997})}\BibitemShut {NoStop}%
\bibitem [{\citenamefont {Mainardi}\ and\ \citenamefont
  {Pironi}(1996)}]{mainardi2008fractional}%
  \BibitemOpen
  \bibfield  {author} {\bibinfo {author} {\bibfnamefont {F.}~\bibnamefont
  {Mainardi}}\ and\ \bibinfo {author} {\bibfnamefont {P.}~\bibnamefont
  {Pironi}},\ }\href@noop {} {\bibfield  {journal} {\bibinfo  {journal}
  {Extracta Math.}\ }\textbf {\bibinfo {volume} {11}},\ \bibinfo {pages} {140}
  (\bibinfo {year} {1996})}\BibitemShut {NoStop}%
\bibitem [{\citenamefont {Mainardi}(1997)}]{MR1611587}%
  \BibitemOpen
  \bibfield  {author} {\bibinfo {author} {\bibfnamefont {F.}~\bibnamefont
  {Mainardi}},\ }\href {\doibase 10.1007/978-3-7091-2664-6\_7} {\ \bibinfo
  {series} {CISM Courses and Lect.},\ \textbf {\bibinfo {volume} {378}},\
  \bibinfo {pages} {291} (\bibinfo {year} {1997})}\BibitemShut {NoStop}%
\bibitem [{\citenamefont {Guo}\ \emph {et~al.}(2013)\citenamefont {Guo},
  \citenamefont {Zeng}, \citenamefont {Li},\ and\ \citenamefont
  {Chen}}]{guo2013numerics}%
  \BibitemOpen
  \bibfield  {author} {\bibinfo {author} {\bibfnamefont {P.}~\bibnamefont
  {Guo}}, \bibinfo {author} {\bibfnamefont {C.}~\bibnamefont {Zeng}}, \bibinfo
  {author} {\bibfnamefont {C.}~\bibnamefont {Li}}, \ and\ \bibinfo {author}
  {\bibfnamefont {Y.}~\bibnamefont {Chen}},\ }\href@noop {} {\bibfield
  {journal} {\bibinfo  {journal} {Fractional Calculus and Applied Analysis}\
  }\textbf {\bibinfo {volume} {16}},\ \bibinfo {pages} {123} (\bibinfo {year}
  {2013})}\BibitemShut {NoStop}%
\bibitem [{\citenamefont {Li}\ and\ \citenamefont {Zeng}(2013)}]{li2013finite}%
  \BibitemOpen
  \bibfield  {author} {\bibinfo {author} {\bibfnamefont {C.}~\bibnamefont
  {Li}}\ and\ \bibinfo {author} {\bibfnamefont {F.}~\bibnamefont {Zeng}},\
  }\href@noop {} {\bibfield  {journal} {\bibinfo  {journal} {Numerical
  Functional Analysis and Optimization}\ }\textbf {\bibinfo {volume} {34}},\
  \bibinfo {pages} {149} (\bibinfo {year} {2013})}\BibitemShut {NoStop}%
\bibitem [{\citenamefont {Miller}\ and\ \citenamefont
  {Ross}(1993)}]{miller1993introduction}%
  \BibitemOpen
  \bibfield  {author} {\bibinfo {author} {\bibfnamefont {K.~S.}\ \bibnamefont
  {Miller}}\ and\ \bibinfo {author} {\bibfnamefont {B.}~\bibnamefont {Ross}},\
  }\href@noop {} {\bibfield  {journal} {\bibinfo  {journal} {John Wiley \&
  Sons, Inc., New York}\ ,\ \bibinfo {pages} {xvi+366}} (\bibinfo {year}
  {1993})}\BibitemShut {NoStop}%
\bibitem [{\citenamefont {Carpinteri}\ and\ \citenamefont
  {Mainardi}(2014)}]{carpinteri2014fractals}%
  \BibitemOpen
  \bibfield  {author} {\bibinfo {author} {\bibfnamefont {A.}~\bibnamefont
  {Carpinteri}}\ and\ \bibinfo {author} {\bibfnamefont {F.}~\bibnamefont
  {Mainardi}},\ }\href@noop {} {\bibfield  {journal} {\bibinfo  {journal}
  {Fractals and Fractional Calculus in Continuum Mechanics}\ }\textbf {\bibinfo
  {volume} {378}} (\bibinfo {year} {2014})}\BibitemShut {NoStop}%
\bibitem [{\citenamefont {Agrawal}(2007)}]{Agrawal_2007}%
  \BibitemOpen
  \bibfield  {author} {\bibinfo {author} {\bibfnamefont {O.~P.}\ \bibnamefont
  {Agrawal}},\ }\href {\doibase 10.1088/1751-8113/40/24/003} {\bibfield
  {journal} {\bibinfo  {journal} {Journal of Physics A: Mathematical and
  Theoretical}\ }\textbf {\bibinfo {volume} {40}},\ \bibinfo {pages} {6287}
  (\bibinfo {year} {2007})}\BibitemShut {NoStop}%
\bibitem [{\citenamefont {Feller}(1971)}]{Feller1971}%
  \BibitemOpen
  \bibfield  {author} {\bibinfo {author} {\bibfnamefont {W.}~\bibnamefont
  {Feller}},\ }\href {https://books.google.co.in/books?id=uPKXwAEACAAJ}
  {\bibfield  {journal} {\bibinfo  {journal} {John Wiley and Sons}\ } (\bibinfo
  {year} {1971})}\BibitemShut {NoStop}%
\bibitem [{\citenamefont {Metzler}\ \emph {et~al.}(1999)\citenamefont
  {Metzler}, \citenamefont {Barkai},\ and\ \citenamefont
  {Klafter}}]{PhysRevLett.82.3563}%
  \BibitemOpen
  \bibfield  {author} {\bibinfo {author} {\bibfnamefont {R.}~\bibnamefont
  {Metzler}}, \bibinfo {author} {\bibfnamefont {E.}~\bibnamefont {Barkai}}, \
  and\ \bibinfo {author} {\bibfnamefont {J.}~\bibnamefont {Klafter}},\ }\href
  {\doibase 10.1103/PhysRevLett.82.3563} {\bibfield  {journal} {\bibinfo
  {journal} {Phys. Rev. Lett.}\ }\textbf {\bibinfo {volume} {82}},\ \bibinfo
  {pages} {3563} (\bibinfo {year} {1999})}\BibitemShut {NoStop}%
\bibitem [{\citenamefont {Metzler}(2000)}]{PhysRevE.62.6233}%
  \BibitemOpen
  \bibfield  {author} {\bibinfo {author} {\bibfnamefont {R.}~\bibnamefont
  {Metzler}},\ }\href {\doibase 10.1103/PhysRevE.62.6233} {\bibfield  {journal}
  {\bibinfo  {journal} {Phys. Rev. E}\ }\textbf {\bibinfo {volume} {62}},\
  \bibinfo {pages} {6233} (\bibinfo {year} {2000})}\BibitemShut {NoStop}%
\bibitem [{\citenamefont {Zumofen}\ and\ \citenamefont
  {Klafter}(1994)}]{zumofen1994spectral}%
  \BibitemOpen
  \bibfield  {author} {\bibinfo {author} {\bibfnamefont {G.}~\bibnamefont
  {Zumofen}}\ and\ \bibinfo {author} {\bibfnamefont {J.}~\bibnamefont
  {Klafter}},\ }\href@noop {} {\bibfield  {journal} {\bibinfo  {journal}
  {Chemical physics letters}\ }\textbf {\bibinfo {volume} {219}},\ \bibinfo
  {pages} {303} (\bibinfo {year} {1994})}\BibitemShut {NoStop}%
\bibitem [{\citenamefont {Lomholt}\ \emph {et~al.}(2005)\citenamefont
  {Lomholt}, \citenamefont {Ambj\"ornsson},\ and\ \citenamefont
  {Metzler}}]{PhysRevLett.95.260603}%
  \BibitemOpen
  \bibfield  {author} {\bibinfo {author} {\bibfnamefont {M.~A.}\ \bibnamefont
  {Lomholt}}, \bibinfo {author} {\bibfnamefont {T.}~\bibnamefont
  {Ambj\"ornsson}}, \ and\ \bibinfo {author} {\bibfnamefont {R.}~\bibnamefont
  {Metzler}},\ }\href {\doibase 10.1103/PhysRevLett.95.260603} {\bibfield
  {journal} {\bibinfo  {journal} {Phys. Rev. Lett.}\ }\textbf {\bibinfo
  {volume} {95}},\ \bibinfo {pages} {260603} (\bibinfo {year}
  {2005})}\BibitemShut {NoStop}%
\bibitem [{\citenamefont {Shlesinger}\ \emph {et~al.}(1993)\citenamefont
  {Shlesinger}, \citenamefont {Zaslavsky},\ and\ \citenamefont
  {Klafter}}]{Shlesinger1993}%
  \BibitemOpen
  \bibfield  {author} {\bibinfo {author} {\bibfnamefont {M.~F.}\ \bibnamefont
  {Shlesinger}}, \bibinfo {author} {\bibfnamefont {G.~M.}\ \bibnamefont
  {Zaslavsky}}, \ and\ \bibinfo {author} {\bibfnamefont {J.}~\bibnamefont
  {Klafter}},\ }\href {\doibase 10.1038/363031a0} {\bibfield  {journal}
  {\bibinfo  {journal} {Nature}\ }\textbf {\bibinfo {volume} {363}},\ \bibinfo
  {pages} {31} (\bibinfo {year} {1993})}\BibitemShut {NoStop}%
\bibitem [{\citenamefont {Wang}\ and\ \citenamefont
  {Tokuyama}(1999)}]{wang1999nonequilibrium}%
  \BibitemOpen
  \bibfield  {author} {\bibinfo {author} {\bibfnamefont {K.}~\bibnamefont
  {Wang}}\ and\ \bibinfo {author} {\bibfnamefont {M.}~\bibnamefont
  {Tokuyama}},\ }\href@noop {} {\bibfield  {journal} {\bibinfo  {journal}
  {Physica A: Statistical Mechanics and its Applications}\ }\textbf {\bibinfo
  {volume} {265}},\ \bibinfo {pages} {341} (\bibinfo {year}
  {1999})}\BibitemShut {NoStop}%
\bibitem [{\citenamefont {Wang}(1992{\natexlab{a}})}]{wang1992long}%
  \BibitemOpen
  \bibfield  {author} {\bibinfo {author} {\bibfnamefont {K.}~\bibnamefont
  {Wang}},\ }\href {\doibase 10.1103/PhysRevA.45.833} {\bibfield  {journal}
  {\bibinfo  {journal} {Physical Review A}\ }\textbf {\bibinfo {volume} {45}},\
  \bibinfo {pages} {833} (\bibinfo {year} {1992}{\natexlab{a}})}\BibitemShut
  {NoStop}%
\bibitem [{\citenamefont {Porr\`a}\ \emph {et~al.}(1996)\citenamefont
  {Porr\`a}, \citenamefont {Wang},\ and\ \citenamefont
  {Masoliver}}]{PhysRevE.53.5872}%
  \BibitemOpen
  \bibfield  {author} {\bibinfo {author} {\bibfnamefont {J.~M.}\ \bibnamefont
  {Porr\`a}}, \bibinfo {author} {\bibfnamefont {K.-G.}\ \bibnamefont {Wang}}, \
  and\ \bibinfo {author} {\bibfnamefont {J.}~\bibnamefont {Masoliver}},\ }\href
  {\doibase 10.1103/PhysRevE.53.5872} {\bibfield  {journal} {\bibinfo
  {journal} {Phys. Rev. E}\ }\textbf {\bibinfo {volume} {53}},\ \bibinfo
  {pages} {5872} (\bibinfo {year} {1996})}\BibitemShut {NoStop}%
\bibitem [{\citenamefont {Wang}(1992{\natexlab{b}})}]{PhysRevA.45.833}%
  \BibitemOpen
  \bibfield  {author} {\bibinfo {author} {\bibfnamefont {K.~G.}\ \bibnamefont
  {Wang}},\ }\href {\doibase 10.1103/PhysRevA.45.833} {\bibfield  {journal}
  {\bibinfo  {journal} {Phys. Rev. A}\ }\textbf {\bibinfo {volume} {45}},\
  \bibinfo {pages} {833} (\bibinfo {year} {1992}{\natexlab{b}})}\BibitemShut
  {NoStop}%
\bibitem [{\citenamefont {Fa}(2006)}]{PhysRevE.73.061104}%
  \BibitemOpen
  \bibfield  {author} {\bibinfo {author} {\bibfnamefont {K.~S.}\ \bibnamefont
  {Fa}},\ }\href {\doibase 10.1103/PhysRevE.73.061104} {\bibfield  {journal}
  {\bibinfo  {journal} {Phys. Rev. E}\ }\textbf {\bibinfo {volume} {73}},\
  \bibinfo {pages} {061104} (\bibinfo {year} {2006})}\BibitemShut {NoStop}%
\bibitem [{\citenamefont {Siegle}\ \emph {et~al.}(2010)\citenamefont {Siegle},
  \citenamefont {Goychuk},\ and\ \citenamefont
  {H\"anggi}}]{PhysRevLett.105.100602}%
  \BibitemOpen
  \bibfield  {author} {\bibinfo {author} {\bibfnamefont {P.}~\bibnamefont
  {Siegle}}, \bibinfo {author} {\bibfnamefont {I.}~\bibnamefont {Goychuk}}, \
  and\ \bibinfo {author} {\bibfnamefont {P.}~\bibnamefont {H\"anggi}},\ }\href
  {\doibase 10.1103/PhysRevLett.105.100602} {\bibfield  {journal} {\bibinfo
  {journal} {Phys. Rev. Lett.}\ }\textbf {\bibinfo {volume} {105}},\ \bibinfo
  {pages} {100602} (\bibinfo {year} {2010})}\BibitemShut {NoStop}%
\bibitem [{\citenamefont {Chen}\ \emph {et~al.}(2023)\citenamefont {Chen},
  \citenamefont {Greiner},\ and\ \citenamefont {Xu}}]{PhysRevE.107.064131}%
  \BibitemOpen
  \bibfield  {author} {\bibinfo {author} {\bibfnamefont {W.}~\bibnamefont
  {Chen}}, \bibinfo {author} {\bibfnamefont {C.}~\bibnamefont {Greiner}}, \
  and\ \bibinfo {author} {\bibfnamefont {Z.}~\bibnamefont {Xu}},\ }\href
  {\doibase 10.1103/PhysRevE.107.064131} {\bibfield  {journal} {\bibinfo
  {journal} {Phys. Rev. E}\ }\textbf {\bibinfo {volume} {107}},\ \bibinfo
  {pages} {064131} (\bibinfo {year} {2023})}\BibitemShut {NoStop}%
\bibitem [{\citenamefont {Hammelmann}\ \emph {et~al.}(2019)\citenamefont
  {Hammelmann}, \citenamefont {Torres-Rincon}, \citenamefont {Rose},
  \citenamefont {Greif},\ and\ \citenamefont {Elfner}}]{PhysRevD.99.076015}%
  \BibitemOpen
  \bibfield  {author} {\bibinfo {author} {\bibfnamefont {J.}~\bibnamefont
  {Hammelmann}}, \bibinfo {author} {\bibfnamefont {J.~M.}\ \bibnamefont
  {Torres-Rincon}}, \bibinfo {author} {\bibfnamefont {J.-B.}\ \bibnamefont
  {Rose}}, \bibinfo {author} {\bibfnamefont {M.}~\bibnamefont {Greif}}, \ and\
  \bibinfo {author} {\bibfnamefont {H.}~\bibnamefont {Elfner}},\ }\href
  {\doibase 10.1103/PhysRevD.99.076015} {\bibfield  {journal} {\bibinfo
  {journal} {Phys. Rev. D}\ }\textbf {\bibinfo {volume} {99}},\ \bibinfo
  {pages} {076015} (\bibinfo {year} {2019})}\BibitemShut {NoStop}%
\bibitem [{\citenamefont {Kapusta}\ \emph {et~al.}(2012)\citenamefont
  {Kapusta}, \citenamefont {M\"uller},\ and\ \citenamefont
  {Stephanov}}]{PhysRevC.85.054906}%
  \BibitemOpen
  \bibfield  {author} {\bibinfo {author} {\bibfnamefont {J.~I.}\ \bibnamefont
  {Kapusta}}, \bibinfo {author} {\bibfnamefont {B.}~\bibnamefont {M\"uller}}, \
  and\ \bibinfo {author} {\bibfnamefont {M.}~\bibnamefont {Stephanov}},\ }\href
  {\doibase 10.1103/PhysRevC.85.054906} {\bibfield  {journal} {\bibinfo
  {journal} {Phys. Rev. C}\ }\textbf {\bibinfo {volume} {85}},\ \bibinfo
  {pages} {054906} (\bibinfo {year} {2012})}\BibitemShut {NoStop}%
\bibitem [{\citenamefont {Ruggieri}\ \emph {et~al.}(2022)\citenamefont
  {Ruggieri}, \citenamefont {Pooja}, \citenamefont {Prakash},\ and\
  \citenamefont {Das}}]{Ruggieri:2022kxv}%
  \BibitemOpen
  \bibfield  {author} {\bibinfo {author} {\bibfnamefont {M.}~\bibnamefont
  {Ruggieri}}, \bibinfo {author} {\bibnamefont {Pooja}}, \bibinfo {author}
  {\bibfnamefont {J.}~\bibnamefont {Prakash}}, \ and\ \bibinfo {author}
  {\bibfnamefont {S.~K.}\ \bibnamefont {Das}},\ }\href {\doibase
  10.1103/PhysRevD.106.034032} {\bibfield  {journal} {\bibinfo  {journal}
  {Phys. Rev. D}\ }\textbf {\bibinfo {volume} {106}},\ \bibinfo {pages}
  {034032} (\bibinfo {year} {2022})},\ \Eprint
  {http://arxiv.org/abs/2203.06712} {arXiv:2203.06712 [hep-ph]} \BibitemShut
  {NoStop}%
\bibitem [{\citenamefont {Pooja}\ \emph {et~al.}(2023)\citenamefont {Pooja},
  \citenamefont {Das}, \citenamefont {Greco},\ and\ \citenamefont
  {Ruggieri}}]{Pooja:2023gqt}%
  \BibitemOpen
  \bibfield  {author} {\bibinfo {author} {\bibnamefont {Pooja}}, \bibinfo
  {author} {\bibfnamefont {S.~K.}\ \bibnamefont {Das}}, \bibinfo {author}
  {\bibfnamefont {V.}~\bibnamefont {Greco}}, \ and\ \bibinfo {author}
  {\bibfnamefont {M.}~\bibnamefont {Ruggieri}},\ }\href {\doibase
  10.1103/PhysRevD.108.054026} {\bibfield  {journal} {\bibinfo  {journal}
  {Phys. Rev. D}\ }\textbf {\bibinfo {volume} {108}},\ \bibinfo {pages}
  {054026} (\bibinfo {year} {2023})},\ \Eprint
  {http://arxiv.org/abs/2306.13749} {arXiv:2306.13749 [hep-ph]} \BibitemShut
  {NoStop}%
\bibitem [{\citenamefont {Liu}\ \emph {et~al.}(2021)\citenamefont {Liu},
  \citenamefont {Das}, \citenamefont {Greco},\ and\ \citenamefont
  {Ruggieri}}]{PhysRevD.103.034029}%
  \BibitemOpen
  \bibfield  {author} {\bibinfo {author} {\bibfnamefont {J.-H.}\ \bibnamefont
  {Liu}}, \bibinfo {author} {\bibfnamefont {S.~K.}\ \bibnamefont {Das}},
  \bibinfo {author} {\bibfnamefont {V.}~\bibnamefont {Greco}}, \ and\ \bibinfo
  {author} {\bibfnamefont {M.}~\bibnamefont {Ruggieri}},\ }\href {\doibase
  10.1103/PhysRevD.103.034029} {\bibfield  {journal} {\bibinfo  {journal}
  {Phys. Rev. D}\ }\textbf {\bibinfo {volume} {103}},\ \bibinfo {pages}
  {034029} (\bibinfo {year} {2021})}\BibitemShut {NoStop}%
\bibitem [{\citenamefont {Caucal}\ and\ \citenamefont
  {Mehtar-Tani}(2022)}]{Caucal:2021lgf}%
  \BibitemOpen
  \bibfield  {author} {\bibinfo {author} {\bibfnamefont {P.}~\bibnamefont
  {Caucal}}\ and\ \bibinfo {author} {\bibfnamefont {Y.}~\bibnamefont
  {Mehtar-Tani}},\ }\href {\doibase 10.1103/PhysRevD.106.L051501} {\bibfield
  {journal} {\bibinfo  {journal} {Phys. Rev. D}\ }\textbf {\bibinfo {volume}
  {106}},\ \bibinfo {pages} {L051501} (\bibinfo {year} {2022})},\ \Eprint
  {http://arxiv.org/abs/2109.12041} {arXiv:2109.12041 [hep-ph]} \BibitemShut
  {NoStop}%
\bibitem [{\citenamefont {Caputo}(1967)}]{10.1111/j.1365-246X.1967.tb02303.x}%
  \BibitemOpen
  \bibfield  {author} {\bibinfo {author} {\bibfnamefont {M.}~\bibnamefont
  {Caputo}},\ }\href {\doibase 10.1111/j.1365-246X.1967.tb02303.x} {\bibfield
  {journal} {\bibinfo  {journal} {Geophysical Journal International}\ }\textbf
  {\bibinfo {volume} {13}},\ \bibinfo {pages} {529} (\bibinfo {year}
  {1967})}\BibitemShut {NoStop}%
\bibitem [{\citenamefont {Podlubny}()}]{podlubnyacademic}%
  \BibitemOpen
  \bibfield  {author} {\bibinfo {author} {\bibfnamefont {I.}~\bibnamefont
  {Podlubny}},\ }\href@noop {} {\bibinfo  {journal} {Fractional Differential
  Equations, Aca- demic Press, San Diego, 1999}\ }\BibitemShut {NoStop}%
\bibitem [{\citenamefont {Sandev}\ \emph {et~al.}(2012)\citenamefont {Sandev},
  \citenamefont {Metzler},\ and\ \citenamefont {Tomovski}}]{Sandev2012}%
  \BibitemOpen
\bibfield  {journal} {  }\bibfield  {author} {\bibinfo {author} {\bibfnamefont
  {T.}~\bibnamefont {Sandev}}, \bibinfo {author} {\bibfnamefont
  {R.}~\bibnamefont {Metzler}}, \ and\ \bibinfo {author} {\bibfnamefont
  {Å.}~\bibnamefont {Tomovski}},\ }\href@noop {} {\bibfield  {journal}
  {\bibinfo  {journal} {Fractional Calculus and Applied Analysis}\ }\textbf
  {\bibinfo {volume} {15}},\ \bibinfo {pages} {426} (\bibinfo {year}
  {2012})}\BibitemShut {NoStop}%
\bibitem [{\citenamefont {Erd{\'e}lyi}\ \emph {et~al.}(1953)\citenamefont
  {Erd{\'e}lyi}, \citenamefont {Magnus}, \citenamefont {Oberhettinger},\ and\
  \citenamefont {Tricomi}}]{erdelyi1953bateman}%
  \BibitemOpen
  \bibfield  {author} {\bibinfo {author} {\bibfnamefont {A.}~\bibnamefont
  {Erd{\'e}lyi}}, \bibinfo {author} {\bibfnamefont {W.}~\bibnamefont {Magnus}},
  \bibinfo {author} {\bibfnamefont {F.}~\bibnamefont {Oberhettinger}}, \ and\
  \bibinfo {author} {\bibfnamefont {F.}~\bibnamefont {Tricomi}},\ }\href@noop
  {} {\bibfield  {journal} {\bibinfo  {journal} {Higher transcendental
  functions}\ }\textbf {\bibinfo {volume} {2}},\ \bibinfo {pages} {133}
  (\bibinfo {year} {1953})}\BibitemShut {NoStop}%
\bibitem [{\citenamefont {Hilfer}(2000)}]{MR1890104}%
  \BibitemOpen
  \bibfield  {author} {\bibinfo {author} {\bibfnamefont {R.}~\bibnamefont
  {Hilfer}},\ }\href {\doibase 10.1142/9789812817747} {\bibfield  {journal}
  {\bibinfo  {journal} {World Scientific Publishing Co., Inc., River Edge, NJ}\
  ,\ \bibinfo {pages} {viii+463}} (\bibinfo {year} {2000})}\BibitemShut
  {NoStop}%
\bibitem [{\citenamefont {Rogers}(1997)}]{rogers1997arbitrage}%
  \BibitemOpen
  \bibfield  {author} {\bibinfo {author} {\bibfnamefont {L.~C.~G.}\
  \bibnamefont {Rogers}},\ }\href@noop {} {\bibfield  {journal} {\bibinfo
  {journal} {Mathematical finance}\ }\textbf {\bibinfo {volume} {7}},\ \bibinfo
  {pages} {95} (\bibinfo {year} {1997})}\BibitemShut {NoStop}%
\bibitem [{\citenamefont {Oldham}\ and\ \citenamefont
  {Spanier}(1974)}]{MR361633}%
  \BibitemOpen
  \bibfield  {author} {\bibinfo {author} {\bibfnamefont {K.~B.}\ \bibnamefont
  {Oldham}}\ and\ \bibinfo {author} {\bibfnamefont {J.}~\bibnamefont
  {Spanier}},\ }\href@noop {} {\bibfield  {journal} {\bibinfo  {journal}
  {Mathematics in Science and Engineering}\ }\textbf {\bibinfo {volume} {111}}
  (\bibinfo {year} {1974})},\ \Eprint {http://arxiv.org/abs/2105.14296}
  {2105.14296} \BibitemShut {NoStop}%
\bibitem [{\citenamefont {Walton}\ and\ \citenamefont
  {Rafelski}(2000)}]{PhysRevLett.84.31}%
  \BibitemOpen
  \bibfield  {author} {\bibinfo {author} {\bibfnamefont {D.~B.}\ \bibnamefont
  {Walton}}\ and\ \bibinfo {author} {\bibfnamefont {J.}~\bibnamefont
  {Rafelski}},\ }\href {\doibase 10.1103/PhysRevLett.84.31} {\bibfield
  {journal} {\bibinfo  {journal} {Phys. Rev. Lett.}\ }\textbf {\bibinfo
  {volume} {84}},\ \bibinfo {pages} {31} (\bibinfo {year} {2000})}\BibitemShut
  {NoStop}%
\bibitem [{\citenamefont {Moore}\ and\ \citenamefont
  {Teaney}(2005)}]{Moore:2004tg}%
  \BibitemOpen
  \bibfield  {author} {\bibinfo {author} {\bibfnamefont {G.~D.}\ \bibnamefont
  {Moore}}\ and\ \bibinfo {author} {\bibfnamefont {D.}~\bibnamefont {Teaney}},\
  }\href {\doibase 10.1103/PhysRevC.71.064904} {\bibfield  {journal} {\bibinfo
  {journal} {Phys. Rev. C}\ }\textbf {\bibinfo {volume} {71}},\ \bibinfo
  {pages} {064904} (\bibinfo {year} {2005})},\ \Eprint
  {http://arxiv.org/abs/hep-ph/0412346} {arXiv:hep-ph/0412346} \BibitemShut
  {NoStop}%
\bibitem [{\citenamefont {Mazumder}\ \emph {et~al.}(2014)\citenamefont
  {Mazumder}, \citenamefont {Bhattacharyya},\ and\ \citenamefont
  {Alam}}]{Mazumder:2013oaa}%
  \BibitemOpen
  \bibfield  {author} {\bibinfo {author} {\bibfnamefont {S.}~\bibnamefont
  {Mazumder}}, \bibinfo {author} {\bibfnamefont {T.}~\bibnamefont
  {Bhattacharyya}}, \ and\ \bibinfo {author} {\bibfnamefont {J.-e.}\
  \bibnamefont {Alam}},\ }\href {\doibase 10.1103/PhysRevD.89.014002}
  {\bibfield  {journal} {\bibinfo  {journal} {Phys. Rev. D}\ }\textbf {\bibinfo
  {volume} {89}},\ \bibinfo {pages} {014002} (\bibinfo {year} {2014})},\
  \Eprint {http://arxiv.org/abs/1305.6445} {arXiv:1305.6445 [nucl-th]}
  \BibitemShut {NoStop}%
\bibitem [{\citenamefont {Svetitsky}(1988)}]{Svetitsky:1987gq}%
  \BibitemOpen
  \bibfield  {author} {\bibinfo {author} {\bibfnamefont {B.}~\bibnamefont
  {Svetitsky}},\ }\href {\doibase 10.1103/PhysRevD.37.2484} {\bibfield
  {journal} {\bibinfo  {journal} {Phys. Rev. D}\ }\textbf {\bibinfo {volume}
  {37}},\ \bibinfo {pages} {2484} (\bibinfo {year} {1988})}\BibitemShut
  {NoStop}%
\bibitem [{\citenamefont {Cacciari}\ \emph {et~al.}(2005)\citenamefont
  {Cacciari}, \citenamefont {Nason},\ and\ \citenamefont
  {Vogt}}]{cacciari2005qcd}%
  \BibitemOpen
  \bibfield  {author} {\bibinfo {author} {\bibfnamefont {M.}~\bibnamefont
  {Cacciari}}, \bibinfo {author} {\bibfnamefont {P.}~\bibnamefont {Nason}}, \
  and\ \bibinfo {author} {\bibfnamefont {R.}~\bibnamefont {Vogt}},\ }\href
  {\doibase 10.1103/PhysRevLett.95.122001} {\bibfield  {journal} {\bibinfo
  {journal} {Phys. Rev. Lett.}\ }\textbf {\bibinfo {volume} {95}},\ \bibinfo
  {pages} {122001} (\bibinfo {year} {2005})},\ \Eprint
  {http://arxiv.org/abs/hep-ph/0502203} {arXiv:hep-ph/0502203} \BibitemShut
  {NoStop}%
\bibitem [{\citenamefont {Cacciari}\ \emph {et~al.}(2012)\citenamefont
  {Cacciari}, \citenamefont {Frixione}, \citenamefont {Houdeau}, \citenamefont
  {Mangano}, \citenamefont {Nason},\ and\ \citenamefont
  {Ridolfi}}]{Cacciari:2012ny}%
  \BibitemOpen
  \bibfield  {author} {\bibinfo {author} {\bibfnamefont {M.}~\bibnamefont
  {Cacciari}}, \bibinfo {author} {\bibfnamefont {S.}~\bibnamefont {Frixione}},
  \bibinfo {author} {\bibfnamefont {N.}~\bibnamefont {Houdeau}}, \bibinfo
  {author} {\bibfnamefont {M.~L.}\ \bibnamefont {Mangano}}, \bibinfo {author}
  {\bibfnamefont {P.}~\bibnamefont {Nason}}, \ and\ \bibinfo {author}
  {\bibfnamefont {G.}~\bibnamefont {Ridolfi}},\ }\href {\doibase
  10.1007/JHEP10(2012)137} {\bibfield  {journal} {\bibinfo  {journal} {JHEP}\
  }\textbf {\bibinfo {volume} {10}},\ \bibinfo {pages} {137} (\bibinfo {year}
  {2012})},\ \Eprint {http://arxiv.org/abs/1205.6344} {arXiv:1205.6344
  [hep-ph]} \BibitemShut {NoStop}%
\end{thebibliography}%
\end{document}